\documentclass[%
 reprint,
superscriptaddress,
 amsmath,amssymb,
 aps,
]{revtex4-2}

\usepackage{graphicx}
\usepackage{dcolumn}
\usepackage{bm}
\usepackage{appendix}

\usepackage{mathtools,rotating} 

\begin{document}

\preprint{APS/123-QED}

\title{Many-body theory and Gaussian-basis implementation of positron annihilation {$\gamma$} spectra on polyatomic molecules}

\author{S. K. Gregg$^\ddag$} \email{sgregg07@qub.ac.uk}
\author{J. P. Cassidy$^\ddag$}%
\author{A. R. Swann$^\ddag$}
\affiliation{
 Centre for Light-Matter Interactions, School of Mathematics and Physics, Queen's University Belfast, University Road, Belfast BT7 1NN, United Kingdom.
}
\author{J. Hofierka}
 \affiliation{
Theoretische Chemie, Physikalisch-Chemisches Institut, Universität Heidelberg, Heidelberg D-69120, Germany.
}
\author{B. Cunningham}
\author{D. G. Green} 
\email{d.green@qub.ac.uk\\$^\ddag$ SKG, JPC and ARS are joint-first authors.}

\affiliation{
 Centre for Light-Matter Interactions, School of Mathematics and Physics, Queen's University Belfast, University Road, Belfast BT7 1NN, United Kingdom.
}

\date{\today}

\begin{abstract}

Doppler-broadened $\gamma$ spectra for positron annihilation on molecules are calculated using many-body theory. By employing Gaussian bases for the electron and positron wavefunctions, a computable expression that involves a four-centre integral over the two-annihilation-photon momenta is derived for the $\gamma$ spectra in the independent particle model approximation to the annihilation vertex, and implemented in the open-source {\tt EXCITON+} code. 
The influence of electron-positron correlations on the $\gamma$ spectra is examined through \textit{ab initio} treatment of the positron wavefunction, whilst corrections to the annihilation vertex are treated approximately via enhancement factors previously calculated [D. G. Green and G. F. Gribakin, Phys.~Rev.~Lett.~{\bf 114}, 093201  (2015)] exactly for atoms. Calculated $\gamma$ spectra for furan and acetonitrile are presented for annihilation from the positron bound state with electrons of individual molecular orbitals. For such annihilation from the positron-molecule bound state, it is found that the magnitude of the partial contribution to the $\gamma$ spectra from individual molecular orbitals depends not just on the orbital energies, but also on the molecular symmetry, more precisely the relative localisation of the positron and electron densities.

\end{abstract}
                    
\maketitle

\section{\label{sec:introduction} Introduction}
Low-energy positrons can annihilate on atomic and molecular electrons, with the dominant annihilation process producing two detectable $\gamma$ rays whose energies are characteristic of the electron's momentum at the instant of annihilation. Although annihilation predominantly occurs on the valence electron subshells in atoms, or highest-occupied molecular orbitals in molecules, owing to
the positron-nuclear repulsion that dominates the static interaction, 
annihilation on the tightly-bound core electrons contributes significantly and dominates the spectrum at large Doppler energy shifts \cite{Green2015,DGG_corelong}. The annihilation signal thus provides elemental specificity and enables sensitive spectroscopy.
In particular, understanding and deciphering the positron annihilation $\gamma$ spectrum is important to properly interpret positron-based materials sciences techniques such as positron-annihilation-induced Auger electron spectroscopy, an ultrasensitive probe of defects \cite{PhysRevB.20.3566,Puska1994,RMPpossolids2013}, and surfaces and porosity in materials \cite{Hugenschmidt2016}; in astrophysics to decode the strong positron annihilation signal from the galactic centre and potentially provide insight on the molecular decomposition of the interstellar medium \cite{leventhal_galacticpositrons, RevModPhys.83.1001}; and to advance positron and positronium-based PET (positron emission tomography) medical imaging \cite{PETbook,PETnew,JPET,RevModPhys.95.021002}. Time-dependent $\gamma$-spectra provide additional diagnostics of materials and positron cooling in atomic and molecular gases relevant to understanding and developing positron traps and high-energy-resolution beams \cite{DGG_cool, DGG_gamcool, AMOC:2016, AMOC:1997}.

Positron interactions with atoms and molecules are characterised by strong many-body correlations that act over both short and long range \cite{Surko:2005,RevModPhys.82.2557,Green2014,Green2015,Hofierka2022}. When considering annihilation, the correlations can be classified into two types \cite{Green2015,DGG_corelong}: (i) those that dress the positron wavefunction,  as described by the positron-molecule correlation potential (self-energy), including polarisation of the molecular electron cloud, screening of the electron-positron Coulomb interaction, and the important process of virtual-positronium formation; and (ii) those that dress the annihilation vertex that describes short-range electron-positron interactions before annihilation. Both enhance the positron-molecule annihilation rate \cite{Iwata1995,Green2014}, and modify the magnitude and shape of the corresponding $\gamma$ spectra \cite{Dunlop2006,Green2013,Green2010,DGG_molgammashort,Green_2012,Green2015}.
Positrons can also become bound to some molecules via vibrational Feshbach resonances. The presence of a positron-molecule bound state increases the annihilation rate by orders of magnitude. Whilst $\gamma$ spectra have been measured for over 50 molecules \cite{Iwata1997_2}, there have been no accurate calculations of $\gamma$ spectra for molecules which take proper account of electron-positron correlations. 

For atoms, variational calculations have produced results in excellent agreement with experiment for He \cite{VanReeth1996}, but cannot be feasibly extended to larger systems. 
One of us developed a many-body theory approach for noble-gas atoms, giving a near-exact description of positron annihilation in core and valence electrons, and beyond this proposed a method to account for important short-range electron-positron correlations \cite{Green2014, Green2015}. Calculating accurate $\gamma$ spectra for molecules is more challenging because they involve multiple atomic centres. For molecules, an approximate LCAO method \cite{DGG_molgammashort,Green_2012}, B3LYP density-functional theory \cite{Green2010} and the nuclear-orbital-plus-molecular-orbital method \cite{Ikabata2018} have been explored as an approach to the problem. 
The low-energy-plane-wave method has also been employed to calculate $\gamma$ spectra for a large number of molecules \cite{Ma2016}, but the conclusions from this work were found to be erroneous \cite{Green2017_comment}. 
There have been no calculations to date that capture the true many-body nature of the problem.

Here, we present an \textit{ab initio} diagrammatic many-body theory (MBT) approach to the calculation of annihilation $\gamma$ spectra for polyatomic molecules. Specifically, we have developed and implemented a Gaussian-basis approach in our highly-parallelised, open-source positron-molecule many-body theory code {\tt EXCITON+}, which was developed from the electronic structure code {\tt EXCITON} \cite{Patterson2010,Patterson2019,Patterson2020} to include positrons, and has been employed in previous work to produce accurate positron-molecule binding energies \cite{Hofierka2022, ArthurBaidoo2024, Hofierka2024, Cassidy2023} and to describe positron-molecule scattering \cite{Rawlins2023} and positronic bonding \cite{Cassidy2024_2}. 
We derive and present the computable expressions, and demonstrate the approach by calculating $\gamma$ spectra for annihilation from the positron bound state on individual molecular orbitals of molecules which bind the positron, specifically acetonitrile and furan. We consider annihilation from the bound state to avoid the extra complexities of the positron scattering states, whose calculation is challenging, but underway\footnote{For atoms the shape of the spectra changes slowly at low positron energy \cite{Green2015,DGG:2017:ef}. On the other hand, for molecules the positron bound state can be highly anisotropic.}. 

Section \ref{sec:theory} outlines our many-body theory approach to calculating the $\gamma$ spectra which is described in more detail in Appendices \ref{app1} and \ref{app:gammaspectrum}. In Section \ref{sec:results}, we present calculated total $\gamma$ spectra for acetonitrile at three levels of many-body theory and then show contributions to annihilation from individual molecular orbitals in acetonitrile and furan calculated at our most sophisticated level of theory. Section \ref{sec:conclusion} contains a summary of the work.

\section{\label{sec:theory}Many-body theory of annihilation {$\gamma$} spectra using Gaussian bases}
\subsection{Annihilation {$\gamma$} spectra}
\begin{figure*}[!htb]
\includegraphics*[width=0.6\textwidth]{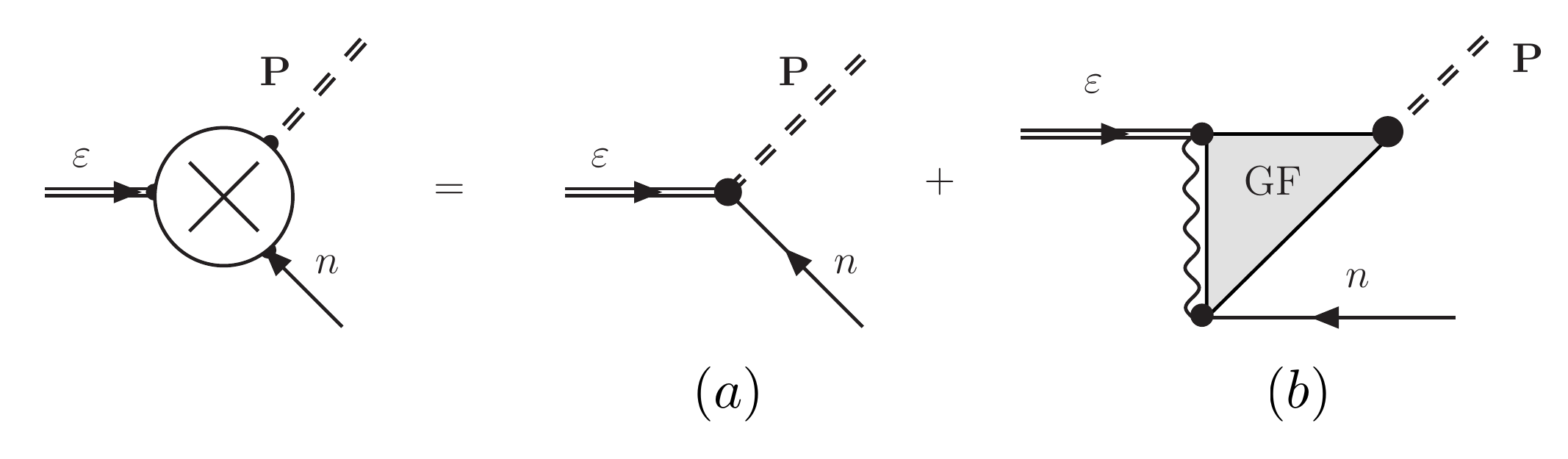}
\caption{Main vertex contributions to the annihilation amplitude $A_{n\varepsilon}({\bf P})$. 
Shown on the right hand side are the different orders of the annihilation vertex: (a) zeroth-order (independent particle approximation); (b) vertex correction in which the positron first excites an electron (thus leaving a hole in state $n$), then undergoes coupled propagation with the excited electron, before they annihilate.
\label{fig:anndiags}}
\end{figure*}

The process of electron-positron annihilation is described fundamentally by 
quantum electrodynamics \cite{Berestetskii1982}. 
It dictates that low-energy positron annihilation proceeds predominately via two photon production, 
a process in which the total spin of the electron-positron system must be zero.
In the centre-of-mass frame, where the total momentum ${\bf P}$ is zero, the two photons 
propagate in opposite directions and have equal energies 
$E_{\gamma}=p_{\gamma}c=mc^2+\frac{1}{2}(E_{\rm i}-E_{\rm f})\simeq mc^2\simeq 511\,{\rm keV}$,
where $E_{\rm i}$ and $E_{\rm f}$ denote the energy of the initial and final states (excluding rest mass). 
When ${\bf P}$ is non-zero however, the two photons no longer propagate in exactly opposite directions
and their energy is Doppler shifted. For example, for the first photon
$E_{\gamma_1}=E_{\gamma}+mcV\cos{\theta}$, where ${\bf V}={\bf P}/2m$ is the centre-of-mass
velocity of the electron-positron pair and $\theta$ is the angle between the direction of the photon velocity ${\bf c}$
and $\bf{V}$.  
Assuming that $V\ll c$, and $p_{\gamma_1}=E_{\gamma_1}/c\approx mc$, the shift of the photon energy from the centre of the line, $\epsilon=E_{\gamma_1}-E_{\gamma}$, is
\begin{eqnarray}
\epsilon=mcV\cos{\theta}=\frac{Pc}{2}\cos{\theta}=\frac{1}{2}{\bf P}\cdot{\bf c}.
\end{eqnarray}
The typical momenta of electrons bound with energy $\varepsilon_n$ determine the Doppler width
 of the annihilation spectrum, $\epsilon \sim Pc \sim\sqrt{|\varepsilon_n|mc^2}\gg|\varepsilon_n|$. 
Hence the shift of the line centre $\varepsilon_n/2$ from $E_\gamma=mc^2=511$\,keV
can usually be neglected, even for the core electrons.
The annihilation $\gamma$-ray (Doppler) spectrum can then be written in a form similar to a Compton profile, 
as the projection of the annihilation probability:
\begin{eqnarray}
w_{n{\bf k}}(\epsilon,{\bf c})=\int |A_{n{\bf k}}({\bf P})|^2 \delta\left(\epsilon-\frac{1}{2}{\bf P}\cdot{\bf c}\right) \frac{d^3P}{(2\pi)^3}.
\end{eqnarray}
For gaseous systems, however, the experimentally accessible quantity is the average of this over 
the direction of emission of the annihilation photons ${\bf c}=c|\hat{\bf c}|$. 
Also, to compare with experiments that use positrons confined in a trap, one must also average over the direction of the incident positron momentum ${\bf k}$ giving 
\begin{eqnarray}\label{eqn:gammaspectra}
\bar{w}_{n\varepsilon}(\epsilon)&\equiv&\int w_{n{\bf k}}(\epsilon,{\bf c}) \frac{d\Omega_{{\bf c}}}{4\pi}\frac{d\Omega_{{\bf k}}}{4\pi},\\
&=&\frac{1}{c} \int _{2|\epsilon|/c}^{\infty} \int_{\Omega_{\bf P}} |A_{n\varepsilon}({\bf P})|^2 \frac{ d\Omega_{\bf P}}{(2\pi)^3} PdP,\label{eqn:gamspec}
\end{eqnarray}
where 
$|A_{n\varepsilon}({\bf P})|^2\equiv\int |A_{n{\bf k}}({\bf P})|^2{d\Omega_{{\bf k}}}/{4\pi}$
is the \emph{annihilation amplitude}. It is the Fourier transform of the correlated electron-positron pair wave function \cite{Green2015,DGG_corelong}.
It can be expanded diagrammatically in the residual electron-electron and electron-positron Coulomb interactions and electron-positron annihilation operator \cite{Green2015,DGG_corelong}.
Figure \ref{fig:anndiags} shows the two main contributions to this expansion.
The first term [Fig.~\ref{fig:anndiags} (a)] is the zeroth-order (IPA) vertex 
\begin{eqnarray}
\label{eq:ann_amp}
A^{(0)}_{n\varepsilon}({\bf P})~& =\int e^{-i{\bf P}\cdot{\bf r}} \psi_{\varepsilon}({\bf r})\varphi_n({\bf r}) d{\bf r},
\end{eqnarray}
i.e., the Fourier transform of the product of the positron $\psi_{\varepsilon}$ and electron $\varphi_n$ wavefunctions.
The second diagram describes the vertex correction (vc) in which the positron first excites an electron (thus leaving a hole in state $n$), before undergoing coupled propagation until they annihilate. It has the abstract form \cite{Green2015,DGG_corelong}
\begin{eqnarray}\label{eqn:annampgeneral}
A^{{\rm vc}}_{n\varepsilon}({\bf P})~& =\int e^{-i{\bf P}\cdot{\bf r}} \tilde\Delta_{\bf P}({\bf r};{\bf r}_1,{\bf r}_2)\psi_{\varepsilon}({\bf r}_1)\varphi_n({\bf r}_2) d{\bf r}_1d{\bf r}_2 d{\bf r},\nonumber\\
\end{eqnarray}
where $\tilde\Delta_{\bf P}$ is the non-local annihilation kernel that describes the coupled propagation.

The annihilation photons need not be detected in coincidence with the final state of the atom. In this case, the observed spectrum is then given by a sum over the electronic states of the system: $\bar{w}_{\varepsilon}(\epsilon)=\sum_n\bar{w}_{n\varepsilon}(\epsilon)$.



\begin{widetext}
\smallskip
\subsection{Calculation of  {$\gamma$} spectrum using Gaussian bases}
\subsubsection{Derivation of computable expression for the zeroth-order $\gamma$ spectrum}
Expanding the positron and electron wavefunctions (molecular orbitals) $\psi_{\varepsilon}(\mathbf{r})$ and $\varphi_n(\mathbf{r})$ in Gaussian bases, 
\begin{align}\label{eq:gaussian_expn}
    \psi_\varepsilon(\mathbf r) &= \sum_i C_i^{\varepsilon} (x\!-\!x_i)^{n^x_i} (y\!-\!y_i)^{n^y_i}(z\!-\!z_i)^{n^z_i} e^{-\zeta_i \lvert \mathbf r-\mathbf r_i\rvert^2},\nonumber\\
\varphi_n(\mathbf r) &=
\sum_j C^{n}_j (x\!-\!x_j)^{n^x_j} (y\!-\!y_j)^{n^y_j}(z\!-\!z_j)^{n^z_j} e^{-\zeta_j \lvert \mathbf r-\mathbf r_j\rvert^2},
\end{align}
where $\mathbf{r}_i = (x_i,y_i,z_i)$ is the centre of the Gaussian function and $n_i^x + n_i^y + n_i^z$ is the angular momentum of the Gaussian,
allows us to derive a computable expression for the zeroth-order $\gamma$ spectrum (see the Appendices for the full derivation), viz.,
\begin{equation}
w^{(0)}_{n \varepsilon}(\epsilon) = 
\sum_{i,i',j,j'}
C^{\varepsilon}_{i}
C^{\varepsilon}_{i'}
C^{n}_{j}
C^{n}_{j'}
[\gamma(\epsilon)]_{i',j'}^{i,j},
\end{equation}
where $[\gamma(\epsilon)]_{i',j'}^{i,j}$ is a 4-centre integral over the total two-$\gamma$ momentum ${\bf P}$ in the atomic basis, the \emph{$\gamma$-spectra matrix}:
\begin{align}\label{eqn:gammat}
&[\gamma(\epsilon)]_{i'j'}^{ij}\equiv
\frac{1}{(2\pi)^3c}\frac{4\pi^4}{[(\zeta_i+\zeta_j)(\zeta_{i'}+\zeta_{j'})]^{3/2}} 
e^{-\lambda_{ij}\lvert \mathbf r_i-\mathbf r_j\rvert^2}
e^{-\lambda_{i'j'}\lvert \mathbf r_{i'}-\mathbf r_{j'}\rvert^2} 
\notag\\
\quad{}&\times \sum_{\substack{s^x_{ij},s^y_{ij},s^z_{ij},\\ s^x_{i'j'},s^y_{i'j'},s^z_{i'j'}}}
E_{s^x_{ij}}^{n^x_i n^x_j} 
E_{s^y_{ij}}^{n^y_i n^y_j} 
E_{s^z_{ij}}^{n^z_i n^z_j} 
E_{s^x_{i'j'}}^{n^x_{i'} n^x_{j'}} 
E_{s^y_{i'j'}}^{n^y_{i'} n^y_{j'}} 
E_{s^z_{i'j'}}^{n^z_{i'} n^z_{j'}}
(-1)^{s^x_{i'j'}+s^y_{i'j'}+s^z_{i'j'}} \notag\\
\quad{}& \times \sum_{\lambda=0}^{s^x_{ij}+s^x_{i'j'}}a^{(s^x_{ij}+s^x_{i'j'})}_\lambda (X_{iji'j'})
 \sum_{\mu=0}^{s^y_{ij}+s^y_{i'j'}} b^{(s^y_{ij}+s^y_{i'j'})}_\mu  (Y_{iji'j'})
 \sum_{\nu=0}^{s^z_{ij}+s^z_{i'j'}}   c^{(s^z_{ij}+s^z_{i'j'})}_\nu (Z_{iji'j'})
\mathfrak I^{(\lambda\mu\nu)}_{iji'j'}(\epsilon) ,
\end{align}
where $\lambda_{ij} = \zeta_i\zeta_j / (\zeta_i + \zeta_j)$, $E$ are the McMurchie-Davidson expansion coefficients \cite{McMurchie1978}, the sums over $s_{ij}^{\mu}$ run from $0$ to $n_{i}^{\mu} + n_j^{\mu}$, $X_{iji'j'}= x_{ij}-x_{i'j'}$ with $x_{ij} = (\zeta_i x_i + \zeta_j x_j)/(\zeta_i + \zeta_j)$, and similarly for $Y_{iji'j'}$ and $Z_{iji'j'}$, the polynomials $a$, $b$ and $c$ are determined using the recursion relations in Table \ref{tab:coeff}  and
\begin{align}
\mathfrak I^{(\lambda\mu\nu)}_{iji'j'}(\epsilon) = 
\int_{2\lvert\epsilon\rvert/c}^\infty P \exp\left[-\frac{P^2}{4}\left(\frac{1}{\zeta_i+\zeta_j}+\frac{1}{\zeta_{i'}+\zeta_{j'}}\right)\right] \mathfrak j_{\lambda+\mu+\nu}(P,R_{iji'j'})\,dP,
\end{align}

where 
\begin{align}
\mathfrak{j}_n(P,R_{iji'j'}) \equiv \left( \frac{P}{R_{iji'j'}}\right)^n j_n(PR_{iji'j'}).
\end{align}
The $j_n(PR_{iji'j'})$ are spherical Bessel functions of order $n$ \cite{Abramowicz} and we note that for zeroth order, $\mathfrak{j}_0(P,R_{iji'j'}) = j_0(P,R_{iji'j'})$.
\end{widetext}
\subsubsection{Annihilation vertex corrections}
We postpone the computationally demanding explicit calculation of the annihilation vertex corrections to future work, instead approximating the effect of the vertex corrections by scaling the zeroth-order $\gamma$-spectra by electron-orbital-specific enhancement factors $\gamma_i$ 
\cite{Green2015},
\begin{equation}\label{enhancement_factor_general}
    \gamma_i = 1 + \sqrt{\frac{A}{-\varepsilon_i}} + \left(\frac{B}{-\varepsilon_i}\right)^{\beta},
\end{equation}
for molecular orbital $i$ with ionisation energy $-\varepsilon_i$ (in eV), where $A$, $B$ and $\beta$ are constants whose values have been determined by fitting to exact calculations for noble-gas atoms \cite{Green2015} and hydrogen-like ions \cite{Green2013} in different approximations to the positron wavefunction \cite{Green2015,DGG:2017:ef} (see below for values used in this work). In addition to accurately describing $\gamma$ spectra for noble-gas atoms \cite{Green2015,DGG:2017:ef}, this enhancement factor formula has been successfully used to calculate the first accurate pickoff annihilation rates for positronium-noble-gas interactions \cite{DGG:2018:PRL,Swann:2023}, and to calculate annihilation of hydrated positronium \cite{Bergami2022}.

\subsubsection{Choice of positron wavefunction: Hartree-Fock vs. Dyson orbital}
In this work we focus on calculating annihilation $\gamma$ spectra for bound-state positrons in polyatomic molecules. The bound-state positron wavefunction is determined by solving the Dyson equation
\begin{equation}\label{Dyson}
    \left(\hat{H}_0 + \hat{\Sigma}_\varepsilon\right)\psi_\varepsilon(\mathbf{r}) = \varepsilon \psi_\varepsilon(\mathbf{r}),
\end{equation}
for the positron-molecule bound state wavefunction $\psi_{\varepsilon}$ and energy $\varepsilon$.
Here $\hat{H}_0$ is the Hamiltonian of the positron in the static (Hartree-Fock) field of the molecule, $\Sigma_{\varepsilon}$ is the positron-molecule self-energy, a nonlocal, energy-dependent potential which accounts for correlations, and $\psi_{\varepsilon}(\mathbf{r})$ is the positron quasiparticle wavefunction with energy $\varepsilon$.
Full details are found in our recent work \cite{Hofierka2022}. Briefly, contributions to the self-energy are calculated in a diagrammatic expansion which involves three infinite classes of diagram: (i) the so-called $GW$ diagram, which describes polarisation of the molecular electron cloud by the positron, screening of the electron-positron Coulomb interaction and electron-hole interactions; (ii) that containing the infinite ladder series of electron-positron interactions that describes the important process of virtual-positronium formation; and (iii) that containing the analogous ladder series describing the repulsive positron-hole interactions. 
The Dyson equation is then solved self-consistently to calculate the positron bound state energy and wavefunction as discussed in Ref. \cite{Hofierka2022}.

The effect of the positron wavefunction correlations on the annihilation $\gamma$ spectra is investigated by calculating the spectra at three levels of theory. At the simplest, Hartree-Fock level of theory the positron-electron interactions are accounted for by considering the positron in the static field of the ground-state molecule in the absence of a correlation potential, i.e., the self-energy operator is set to zero in Equation~\ref{Dyson}. In the second, $GW$@BSE (Bethe-Salpeter equation) level, polarisation, screening of the electron-positron Coulomb interaction and electron-hole attraction are accounted for. The third, full many-body theory approach, $GW{\rm @BSE}+ \tilde\Gamma+\tilde\Lambda$, additionally accounts for virtual positronium formation ($\tilde\Gamma$) and positron-hole repulsion ($\tilde\Lambda$).  

When the Hartree-Fock approximation is used, the enhancement factors $\gamma_i$ of Equation~\ref{enhancement_factor_general} are calculated using $A = 1.544$, $B=0.915$, $\beta=2.54$ and $\varepsilon_i$ are the Hartree-Fock orbital energies.
For the $GW$@BSE and more sophisticated approximations, we instead use $A=1.312$, $B=0.834$, $\beta=2.15$
and $\varepsilon_i$ are the $GW$ orbital energies (those calculated by solving the Dyson equation for the molecule in the diagonal approximation with electron-molecule self-energy calculated at $GW$ level \cite{Hofierka2022}).

\subsubsection{Implementation in  {\tt EXCITON+} and choice of Gaussian bases in present calculations}
We have implemented the above in our open source, Gaussian-basis positron-molecule many-body code, {\tt EXCITON+} \cite{Hofierka2022}. {\tt EXCITON+} is adapted from the all-electron {\tt EXCITON} code \cite{Patterson2010,Patterson2019,Patterson2020}. Both calculate 4-centre Coulomb matrix elements via MPI parallelisation in which each processor handles an individual `quad' (unique combination of the 4 centres). The 4-centre $\gamma$-spectra matrix of Equation~\ref{eqn:gammat} is calculated in the atomic basis via the same parallelisation scheme, before being contracted with the appropriate sets of expansion coefficients. 
We verified the accuracy of the numerical implementation for noble gas atoms and found excellent agreement for calculations at the Hartree-Fock level of theory with the zeroth-order annihilation vertex \cite{Cassidy:thesis}. 

For the calculations presented in this work, Dunning's aug-cc-pVXZ basis sets (X=T, Q) for both the electron and positron wavefunctions \cite{Dunning1992} were placed on each atom in the molecule considered, and also at a number of additional `ghost' centres around the molecule, typically placed at regions of high positron density approximately 1 $\mathrm{\AA}$ from the molecule, to improve the `resolution' (description of electron-positron correlations). A diffuse even-tempered basis for the positron is also placed near the region of high positron density to capture long-range effects; for these calculations, we used an even-tempered basis of the form $10s\,9p\,8d\,7f\,3g$ with exponents $\zeta_0\beta^{k-1}$, $k=1,2,...$, $\zeta_0=0.001$ and $\beta=2.2$.

\section{\label{sec:results}Results and discussion}
We present many-body theory calculated $\gamma$ spectra for annihilation of the positron bound state in acetonitrile (CH$_3$CN) and furan (C$_4$H$_4$O). 
It has been previously established that electron-positron correlations have a crucial influence on the positron-molecule bound state \cite{Hofierka2022} and thus, there is a significant change in the bound state energy and wavefunction obtained at each of the three levels of many-body theory in our investigation. We therefore expect that correlations will also have a substantial influence on the $\gamma$ spectra for annihilation from the positron bound state.

\begin{figure}[t!]
    \centering
    \includegraphics[trim={7.5cm 5cm 7.5cm 5cm},clip,width=0.5\textwidth]{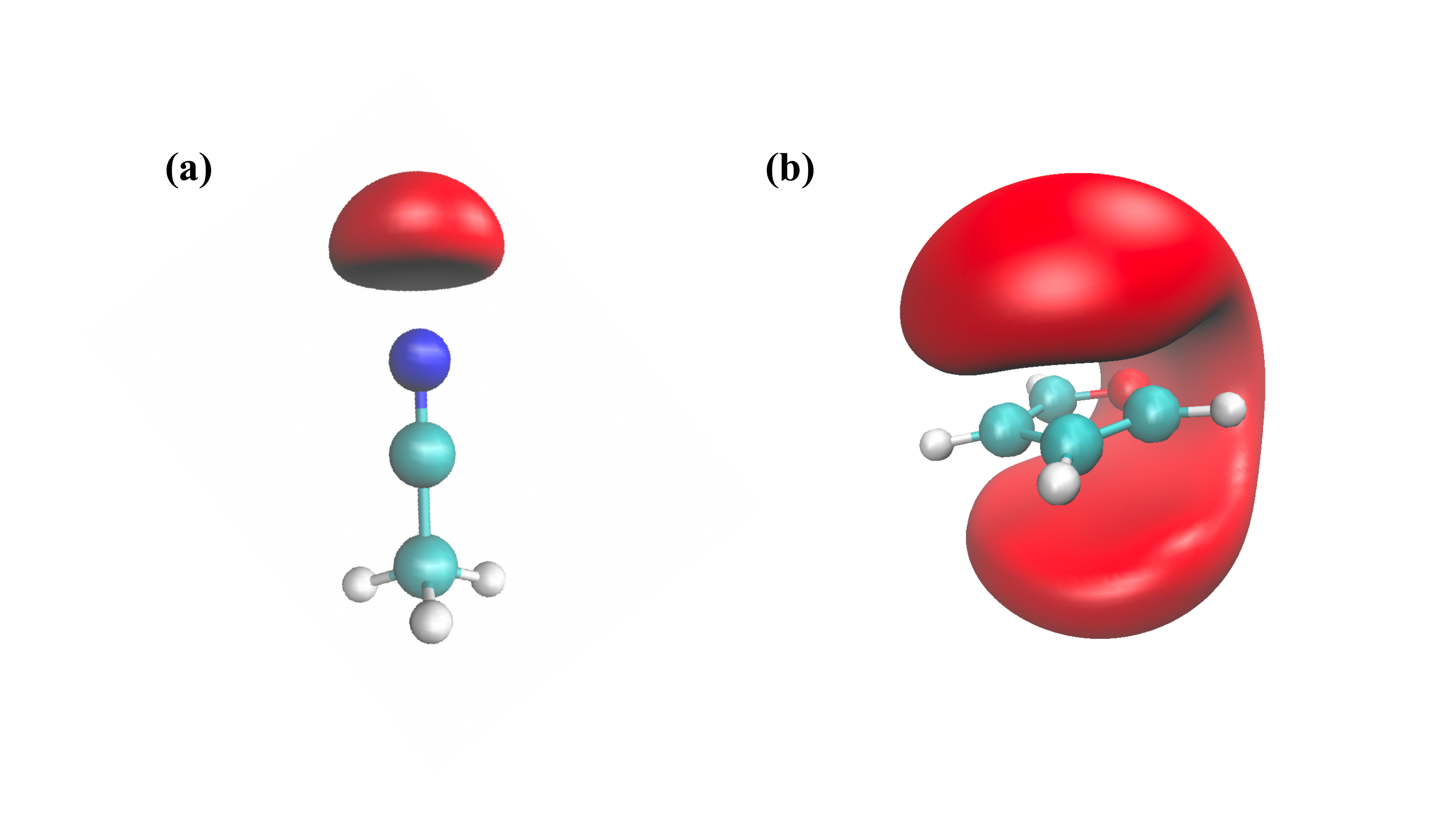}
    \caption{Positron Dyson orbitals for (a) acetonitrile and (b) furan, calculated at the $GW$@BSE $+\tilde\Gamma+\tilde\Lambda$ level of many-body theory. White, green, blue and red spheres are hydrogen, carbon, nitrogen and oxygen atoms, respectively. Solid red isosurfaces are the positron Dyson wavefunctions at 80\%
of their maximum values. }
    \label{fig:dyson_orbitals}
\end{figure}

For acetonitrile, the fully correlated bound state energy for the positron is $\varepsilon_{\rm{b}} =$ 207 meV, the $GW$@BSE binding energy is $\varepsilon_{\rm{b}} =$ 108 meV, and the Hartree-Fock method gives $\varepsilon_{\rm{b}} =$ 11.9 meV. For furan, the fully correlated bound state energy is 42 meV, but no binding is found at the Hartree-Fock or $GW$@BSE levels.
The positron bound-state wavefunctions (or Dyson orbitals) for acetonitrile and furan are shown in Figure \ref{fig:dyson_orbitals}, as calculated at the $GW$@BSE $+ \tilde\Gamma + \tilde\Lambda$ level of many-body theory. 

The $\gamma$ spectra presented in Figures \ref{fig:totals}, \ref{fig:aceto_MOs} and \ref{fig:furan_MOs} of this work were calculated at eleven discrete energy shifts, linearly spaced between 0-10 keV, and interpolated to generate a smooth curve using the cubic spline interpolation function in xmgrace \cite{xmgrace}.

\subsection{Total {$\gamma$} spectra for acetonitrile {(CH$_3$CN)}}
\begin{figure*}[ht]
    \centering
    \includegraphics[trim={2cm 7cm 2cm 5cm},clip,width=\textwidth]{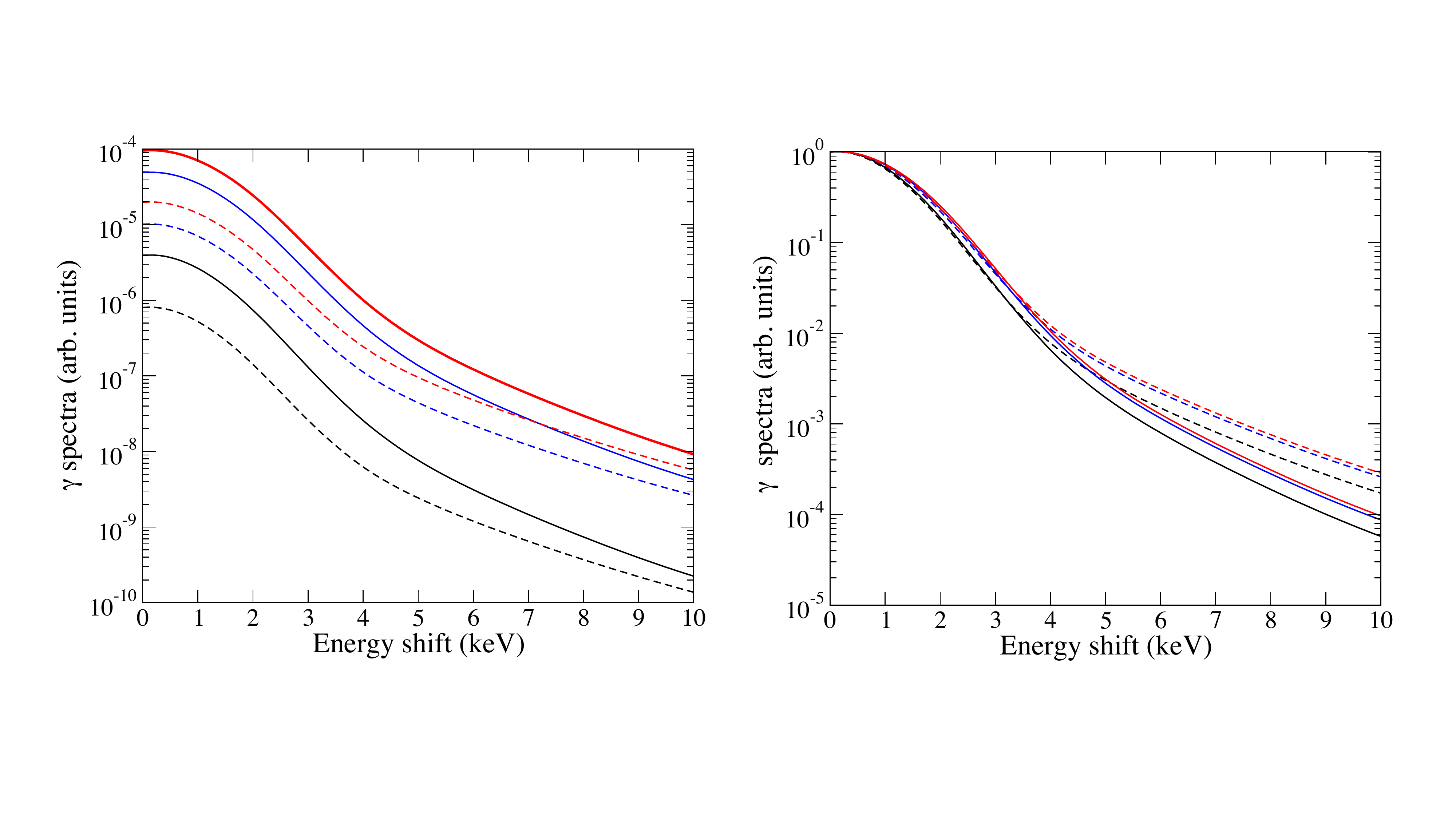}
    \caption{Total $\gamma$ spectra (left) and total $\gamma$ spectra normalised to unity at $\epsilon=$ 0 keV (right) for acetonitrile calculated at three levels of theory: Hartree-Fock (black, $\varepsilon_{\rm b}$ = 11.9 meV), $GW$@BSE (blue, $\varepsilon_{\rm b}$ = 108 meV) and $GW+\tilde\Gamma+\tilde\Lambda$ (red, $\varepsilon_{\rm b}$ = 207 meV). The $\gamma$ spectra are shown both with and without enhancement factors (solid and dashed lines, respectively). }
    \label{fig:totals}
\end{figure*}

Figure \ref{fig:totals} shows calculated total $\gamma$ spectra for annihilation from the bound state of acetonitrile at three levels of many-body theory, both with and without the enhancement factors from Equation \ref{enhancement_factor_general}. Total $\gamma$ spectra normalised to unity at $\epsilon=0$ keV are also shown to highlight changes in the width of the spectra calculated at different levels of many-body theory.  

The positron and electron bases used in these calculations for acetonitrile are identical to those used for the binding calculation in Ref. \cite{Hofierka2022}: aug-cc-pVTZ basis sets on H atoms, aug-cc-pVQZ basis sets on C atoms, aug-cc-pVQZ hydrogen basis sets on the ghost centre, which was placed 1 $\text{\AA}$ away from the N atom along the primary axis of the molecule, and for the N atom, an aug-cc-pVQZ electron basis and an even-tempered positron basis of the form 10$s$9$p$8$d$7$f$3$g$, with smallest exponent $\zeta_0=10^{-3}$ and ratio between exponents $\beta=$ 2.2. For this molecule, a single ghost centre is sufficient to enhance the basis in the region of high positron density, since the positron wavefunction is highly localised to the negative end of the molecule, next to the N atom.

The effect of many-body correlations on the annihilation $\gamma$ spectra is evident from Figure \ref{fig:totals}. Comparing the $\gamma$ spectrum obtained from the Hartree-Fock method to that from the $GW$@BSE level of theory, we find that the magnitude of the spectrum increases approximately tenfold when polarisation and screening of the positron-molecule Coulomb interaction are accounted for. The magnitude of the spectrum is further increased by using the $GW+ \tilde\Gamma+\tilde\Lambda$ self-energy. When the calculated positron bound state has a larger binding energy, the positron wavefunction density is closer to the molecule and thus, overlaps more with the electron orbitals leading to a higher annihilation rate. We note that enhancement factors $\gamma_i$ also increase the magnitude of the total spectra, but to a lesser degree than the wavefunction correlations. 

The normalised total $\gamma$ spectra in Figure \ref{fig:totals} show that the $\gamma$ spectrum is broadened considerably by the inclusion of correlation effects, with the most significant difference seen between the Hartree-Fock and $GW$@BSE spectra. Electron-positron correlations accelerate the positron towards the electron, which increases the momentum distribution and broadens the $\gamma$ spectrum. Enhancement factors have the opposite effect, narrowing the $\gamma$ spectra.

\subsection{Individual molecular orbital contributions to the {$\gamma$} spectra}

In this section, we consider in detail the relative contributions of individual molecular orbitals to the total $\gamma$ spectra for furan and acetonitrile.  

\begin{figure*}[!ht]
    \centering
\includegraphics[trim={0cm 5cm 0cm 0cm},clip,width=\textwidth]{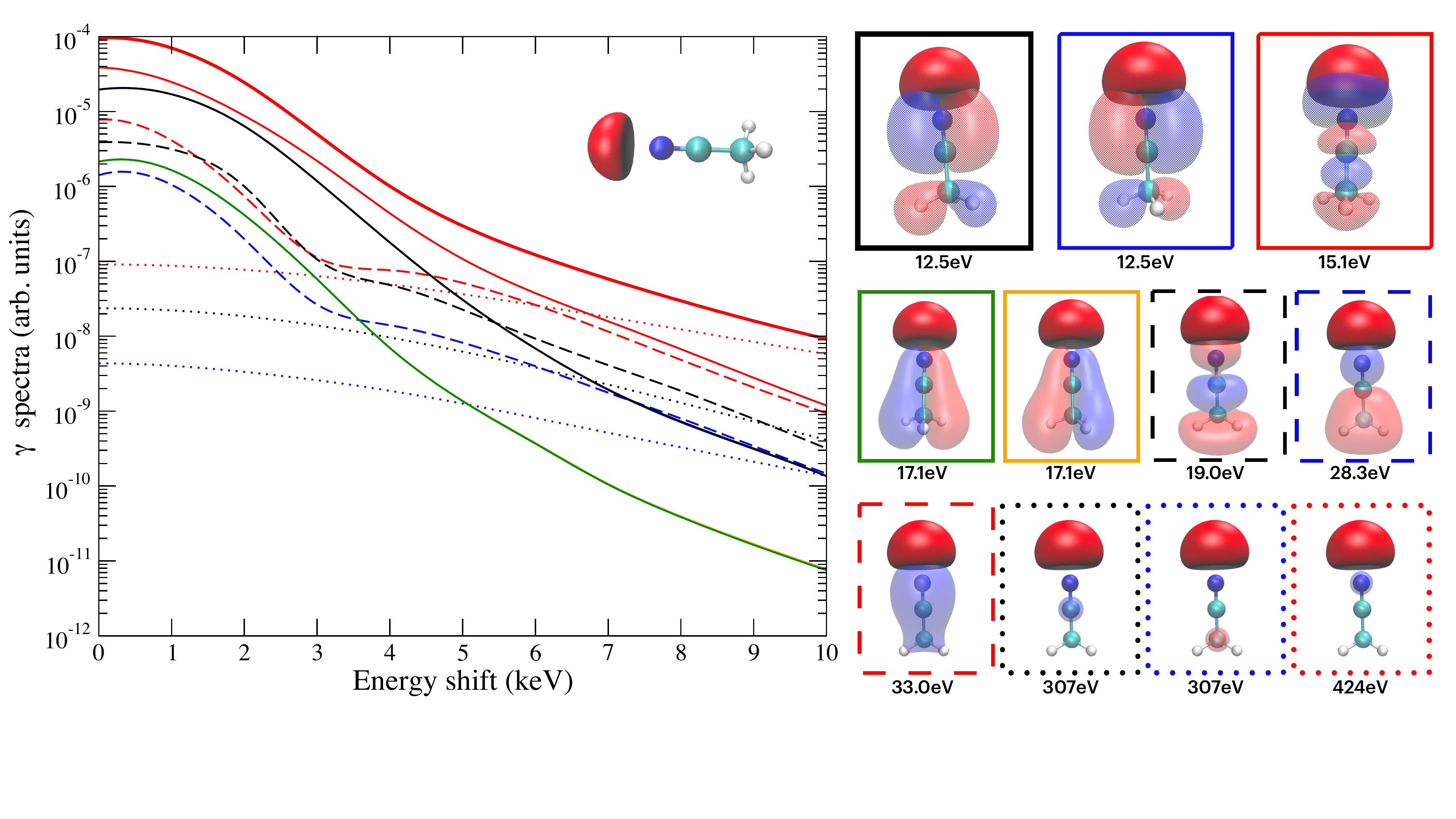}
    \caption{Contributions to $\gamma$ spectra from individual molecular orbitals for acetonitrile (CH$_3$CN) calculated at the $GW\rm{@BSE}+ \tilde\Gamma+\tilde\Lambda$ level of MBT with enhancement factors. The thick red line is the total $\gamma$ spectrum. Electron isosurfaces are shown for each of the molecular orbitals at an isosurface value of $\pm$0.04, labelled with their ionisation energy and with the border of each image matching the curve for its $\gamma$ spectrum.}
    \label{fig:aceto_MOs}
\end{figure*}

\subsubsection{Acetonitrile {(CH$_3$CN)}}
Figure \ref{fig:aceto_MOs} shows contributions to the total $\gamma$ spectra of acetonitrile from each of its 11 molecular orbitals at the $GW$@BSE$+\tilde\Gamma+\tilde\Lambda$ level of theory with enhancement factors (positron binding energy $\varepsilon_{\rm b}$ = 207 meV). Partial annihilation rates for each molecular orbital as a percentage of the total annihilation rate are found in Table \ref{table:annpc}, obtained using the area underneath the $\gamma$ spectrum curves.

It is expected that annihilation rates for core molecular orbitals will be relatively small, since tightly-bound electrons are closer to the repulsive atomic nuclei and thus, are less accessible to the positron for annihilation.  In acetonitrile, three of the 11 molecular orbitals are core orbitals with ionisation energies $\varepsilon_i>$ 300 eV. These describe the most tightly-bound electrons to the N atom and each of the two C atoms, and their combined contribution to annihilation is $<$1\% of the total annihilation rate in the enhanced $GW+\tilde\Gamma+\tilde\Lambda$ annihilation spectrum. Electrons in these core orbitals have high momenta which lead to large energy shifts $\epsilon$ and characteristically broad annihilation spectra.
Relative contributions to annihilation from the three core orbitals in acetonitrile differ by over an order of magnitude, and it is actually the most tightly bound of the three that makes the largest contribution to annihilation. This is understood by noting that the core orbital of the N atom is by far the closest to the region of high positron density of the three and thus, it overlaps significantly more with the positron wavefunction, despite its larger ionisation energy. Contributions to annihilation from the five next most tightly bound orbitals ($\varepsilon_i=$ 33.0-17.1 eV) are also strongly dependent on the locations of their high electron density.

The largest contributions to annihilation come from the three highest occupied molecular orbitals (HOMOs) which combined, are responsible for 83.6\% of the total annihilation rate. However, it is the third most tightly bound orbital which makes the highest individual contribution to annihilation due to its large overlap with the positron orbital (see Figure \ref{fig:aceto_MOs}). Thus, the relative contribution to annihilation from each molecular orbital depends on both the orbital energy and location.

{
\begin{table}[ht!]
\caption{Calculated percentage contributions to the total annihilation rate from the positron-molecule bound state for individual molecular orbitals (ionisation energy $-\varepsilon_i$) in acetonitrile and furan at the $GW$@BSE + $ \tilde\Gamma+\tilde\Lambda$ level of many-body theory with enhancement factors. \label{table:annpc}} 
 \begin{ruledtabular}
 \begin{tabular}{cccc}
\multicolumn{2}{c}{\textbf{Acetonitrile}}  & \multicolumn{2}{c}{\textbf{Furan}}	\\
$-\varepsilon_i$/eV & \% & $-\varepsilon_i$/eV & \% \\

\hline
424		& 	0.28	&561& 		0.05			\\

307     & 	0.01	&307& 		0.07		\\
307     & 	0.06	&307& 		0.07	\\
33.0    & 	6.08	&305& 		0.07		\\
28.2    & 	1.40	&305& 		0.07	\\
19.0    & 	4.20	&40.1& 		2.47	\\
17.1    & 	2.19	&29.6& 		3.31	\\
17.1    & 	2.19	&27.6& 		3.60	\\
15.1    &	36.92	&22.2& 		4.64	\\
12.5    &	23.35	&21.2& 		5.46	\\
12.5    & 	23.34	&20.3& 		5.59	\\
    &		&17.5& 		9.33	\\
    &  		 		 		&16.7&		6.42	\\
    &  		 		 		&15.8& 		5.97	\\    
    &  		 		 		&15.4&		7.72	\\    
    &  		 	 		    &14.6&		7.09	\\
    &  		 				&10.8& 		16.90	\\    
    &  		 				&8.9&		21.13	\\    

\end{tabular}
 \end{ruledtabular}
 \end{table}
} 

\subsubsection{Furan {(C$_4$H$_4$O)}}

\begin{figure*}[!ht]
    \centering
\includegraphics[trim={0cm 5cm 0cm 0},clip,width=\textwidth]{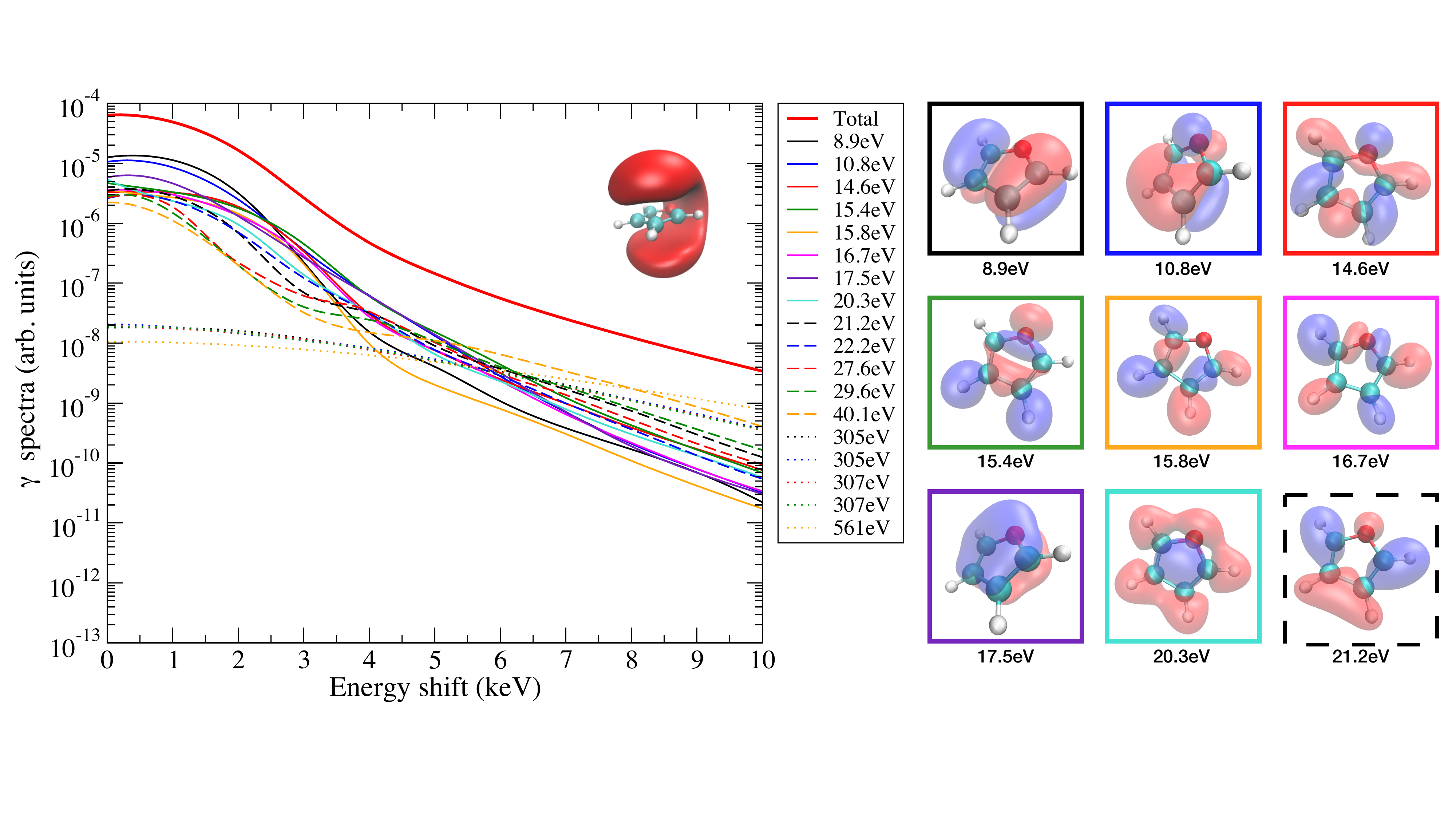}
    \caption{Contributions to $\gamma$ spectra from individual molecular orbitals for furan (C$_4$H$_4$O) calculated at the $GW\rm{@BSE}+ \tilde\Gamma+\tilde\Lambda$ level of MBT with enhancement factors. Electron molecular orbitals are labelled by their binding energy in the graph legend. The thick red line is the total $\gamma$ spectrum. Electron isosurfaces are shown for the nine HOMOs at an isosurface value of $\pm$0.04, labelled with their ionisation energy and with the border of each image matching the curve for its $\gamma$ spectrum.}
    \label{fig:furan_MOs}
\end{figure*}

Figure \ref{fig:furan_MOs} shows contributions to the total $\gamma$ spectra of furan from each of its 18 molecular orbitals at the $GW$@BSE $+\tilde\Gamma+\tilde\Lambda$ level of many-body theory with enhancement factors (binding energy $\varepsilon_{\rm{b}}=$ 42 meV). Partial annihilation rates for each molecular orbital as a percentage of the total annihilation rate are found in Table \ref{table:annpc}.

The positron and electron bases used in this calculation are as follows: aug-cc-pVTZ basis sets on H and C atoms, aug-cc-pVQZ basis sets on the O atom, aug-cc-pVQZ hydrogen basis sets on nine ghost centres placed in the region of high positron density, and for a further ghost centre at the centre of the molecule's ring, an aug-cc-pVQZ hydrogen electron basis and an even-tempered positron basis of the form 10$s$9$p$8$d$7$f$3$g$, with smallest exponent $\zeta_0=10^{-3}$ and ratio between exponents $\beta=$ 2.2.

As for acetonitrile, the $\gamma$ spectra of the tightly-bound core electron orbitals in furan are easily identified by their characteristically broad shape and relatively low levels of contribution to annihilation. From Table \ref{table:annpc}, we find that electrons from the five tightly-bound core molecular orbitals (one for each of the four C atoms and one for the O atom) account for only 0.33\% of the total annihilation rate in the enhanced $GW$@BSE $+\tilde\Gamma+\tilde\Lambda$ spectra. Compared with acetonitrile, we do not see such a pronounced variation in the relative annihilation rates from the core orbitals in furan; this is due to the fact that unlike the core orbitals in acetonitrile, the core orbitals in the furan ring all share similar proximity to the region of high positron density, which is much more delocalised than that for acetonitrile, so their most tightly-bound electrons have a similar probability of annihilation. 

The symmetry of molecular orbitals affects the probability of annihilation with the electrons within them, which can result in more tightly-bound orbitals exhibiting larger contributions to the $\gamma$ spectra. 
In general, molecular orbitals are more accessible to the positron when their electron density is concentrated away from the repulsive atomic centres: they require less tunneling of the positron into the repulsive potential to access them. 
Furan is an aromatic molecule, meaning that it has regions of high electron density above and below the molecular plane due to the presence of $\pi$ orbitals in the ring, which attracts the positron to these regions.
There are three molecular orbitals in furan with $\pi$-type symmetry, and these three $\pi$-type orbitals have the highest partial annihilation rates of all the molecular orbitals in furan. Two of these are also the HOMOs, so their high contribution to annihilation may be attributed to both their ionisation energy and symmetry. However, the third highest contribution comes from the seventh most tightly bound electron orbital ($\varepsilon_i = -17.5$ eV): with regions of high electron density above and below the plane of the molecule, where the positron bound state also has a high density, this molecular orbital has a large overlap with the positron Dyson orbital, resulting in enhanced annihilation and a disruption in the ordering of contributions to annihilation with ionisation energy.

\section{Summary}\label{sec:conclusion}
The many-body theory of Doppler-broadened $\gamma$-spectra for positron annihilation in polyatomic molecules was developed and its implementation via Gaussian bases in the {\tt EXCITON+} code presented. A key result was the derivation of a computable expression involving a 4-centre integral for the $\gamma$ spectrum in the zeroth-order (independent particle approximation) to the annihilation vertex, Equation~\ref{eqn:gammat}, and an efficient parallelised implementation to calculate it.

Calculated $\gamma$ spectra were presented for acetonitrile and furan. Electron-positron correlations were calculated \textit{ab initio} and were found to enhance the  $\gamma$ spectra by 1-2 orders of magnitude and to increase the width of the spectra.
Our calculations were completed using the zeroth-order annihilation vertex, and an established scaling factor was used to recover the effect of vertex corrections on the spectra.
Analysis of individual molecular orbital contributions to the spectra revealed that the largest contributions to the total spectrum tend to come from the least tightly bound orbitals, as expected, but some deviations from the energy-based ordering exist due to the shapes and localisation of the molecular orbitals. In particular, the $\pi$ orbitals in furan make an enhanced contribution to binding, and for the polar molecule acetonitrile, the annihilation rate was much higher for orbitals with high electron density towards the negative end of the molecule.
The latter fact suggests that the energy-dependent enhancement factors may insufficiently capture the effect of the short-range correlations in molecules, and motivates the \emph{ab initio} calculation of the full annihilation amplitude including the vertex correction of Fig.~\ref{fig:anndiags}, which will be the subject of future work, as will calculations of $\gamma$ spectra from annihilation of continuum positrons in molecules.

\section{Author contributions}
S. K. Gregg, J. P. Cassidy and A. R. Swann are joint-first authors. SKG and JPC performed the calculations. 
ARS, JH developed the numerical implementation, supported by BC and DGG. DGG conceived and directed the work. SKG, ARS and DGG drafted the manuscript.

\section{Acknowledgements}
We thank Charles Patterson and Gleb Gribakin for useful discussions. The work was supported by European Research Council Starting Grant 804383 `ANTI-ATOM', and PhD studentships (SKG and JPC) from the Department of Economy, Northern Ireland. We acknowledge use of the UK Tier-2 supercomputer Kelvin2, supported by Northern Ireland High-Performance Computing.

%


\begin{thebibliography}{50}%
\makeatletter
\providecommand \@ifxundefined [1]{%
 \@ifx{#1\undefined}
}%
\providecommand \@ifnum [1]{%
 \ifnum #1\expandafter \@firstoftwo
 \else \expandafter \@secondoftwo
 \fi
}%
\providecommand \@ifx [1]{%
 \ifx #1\expandafter \@firstoftwo
 \else \expandafter \@secondoftwo
 \fi
}%
\providecommand \natexlab [1]{#1}%
\providecommand \enquote  [1]{``#1''}%
\providecommand \bibnamefont  [1]{#1}%
\providecommand \bibfnamefont [1]{#1}%
\providecommand \citenamefont [1]{#1}%
\providecommand \href@noop [0]{\@secondoftwo}%
\providecommand \href [0]{\begingroup \@sanitize@url \@href}%
\providecommand \@href[1]{\@@startlink{#1}\@@href}%
\providecommand \@@href[1]{\endgroup#1\@@endlink}%
\providecommand \@sanitize@url [0]{\catcode `\\12\catcode `\$12\catcode
  `\&12\catcode `\#12\catcode `\^12\catcode `\_12\catcode `\%12\relax}%
\providecommand \@@startlink[1]{}%
\providecommand \@@endlink[0]{}%
\providecommand \url  [0]{\begingroup\@sanitize@url \@url }%
\providecommand \@url [1]{\endgroup\@href {#1}{\urlprefix }}%
\providecommand \urlprefix  [0]{URL }%
\providecommand \Eprint [0]{\href }%
\providecommand \doibase [0]{https://doi.org/}%
\providecommand \selectlanguage [0]{\@gobble}%
\providecommand \bibinfo  [0]{\@secondoftwo}%
\providecommand \bibfield  [0]{\@secondoftwo}%
\providecommand \translation [1]{[#1]}%
\providecommand \BibitemOpen [0]{}%
\providecommand \bibitemStop [0]{}%
\providecommand \bibitemNoStop [0]{.\EOS\space}%
\providecommand \EOS [0]{\spacefactor3000\relax}%
\providecommand \BibitemShut  [1]{\csname bibitem#1\endcsname}%
\let\auto@bib@innerbib\@empty
\bibitem [{\citenamefont {Green}\ and\ \citenamefont
  {Gribakin}(2015)}]{Green2015}%
  \BibitemOpen
  \bibfield  {author} {\bibinfo {author} {\bibfnamefont {D.~G.}\ \bibnamefont
  {Green}}\ and\ \bibinfo {author} {\bibfnamefont {G.~F.}\ \bibnamefont
  {Gribakin}},\ }\bibfield  {title} {\bibinfo {title} {$\ensuremath{\gamma}$
  spectra and enhancement factors for positron annihilation with core
  electrons},\ }\href {https://doi.org/10.1103/PhysRevLett.114.093201}
  {\bibfield  {journal} {\bibinfo  {journal} {Phys. Rev. Lett.}\ }\textbf
  {\bibinfo {volume} {114}},\ \bibinfo {pages} {093201} (\bibinfo {year}
  {2015})}\BibitemShut {NoStop}%
\bibitem [{\citenamefont {Green}\ and\ \citenamefont
  {Gribakin}()}]{DGG_corelong}%
  \BibitemOpen
  \bibfield  {author} {\bibinfo {author} {\bibfnamefont {D.~G.}\ \bibnamefont
  {Green}}\ and\ \bibinfo {author} {\bibfnamefont {G.~F.}\ \bibnamefont
  {Gribakin}},\ }\href@noop {} {\bibinfo {title} {Positron annihilation on core
  and valence electrons}},\ \Eprint {https://arxiv.org/abs/arXiv:1502.08045}
  {arXiv:1502.08045} \BibitemShut {NoStop}%
\bibitem [{\citenamefont {Lynn}\ \emph {et~al.}(1979)\citenamefont {Lynn},
  \citenamefont {Dickman}, \citenamefont {Brown}, \citenamefont {Robbins},\
  and\ \citenamefont {Bonderup}}]{PhysRevB.20.3566}%
  \BibitemOpen
  \bibfield  {author} {\bibinfo {author} {\bibfnamefont {K.}~\bibnamefont
  {Lynn}}, \bibinfo {author} {\bibfnamefont {J.}~\bibnamefont {Dickman}},
  \bibinfo {author} {\bibfnamefont {W.}~\bibnamefont {Brown}}, \bibinfo
  {author} {\bibfnamefont {M.}~\bibnamefont {Robbins}},\ and\ \bibinfo {author}
  {\bibfnamefont {E.}~\bibnamefont {Bonderup}},\ }\bibfield  {title} {\bibinfo
  {title} {Vacancies studied by positron annihilation with high-momentum core
  electrons},\ }\href {https://doi.org/10.1103/PhysRevB.20.3566} {\bibfield
  {journal} {\bibinfo  {journal} {Phys. Rev. B}\ }\textbf {\bibinfo {volume}
  {20}},\ \bibinfo {pages} {3566} (\bibinfo {year} {1979})}\BibitemShut
  {NoStop}%
\bibitem [{\citenamefont {Puska}\ and\ \citenamefont
  {Nieminen}(1994)}]{Puska1994}%
  \BibitemOpen
  \bibfield  {author} {\bibinfo {author} {\bibfnamefont {M.~J.}\ \bibnamefont
  {Puska}}\ and\ \bibinfo {author} {\bibfnamefont {R.~M.}\ \bibnamefont
  {Nieminen}},\ }\bibfield  {title} {\bibinfo {title} {Theory of positrons in
  solids and on solid surfaces},\ }\href
  {https://doi.org/10.1103/RevModPhys.66.841} {\bibfield  {journal} {\bibinfo
  {journal} {Rev. Mod. Phys.}\ }\textbf {\bibinfo {volume} {66}},\ \bibinfo
  {pages} {841} (\bibinfo {year} {1994})}\BibitemShut {NoStop}%
\bibitem [{\citenamefont {Tuomisto}\ and\ \citenamefont
  {Makkonen}(2013)}]{RMPpossolids2013}%
  \BibitemOpen
  \bibfield  {author} {\bibinfo {author} {\bibfnamefont {F.}~\bibnamefont
  {Tuomisto}}\ and\ \bibinfo {author} {\bibfnamefont {I.}~\bibnamefont
  {Makkonen}},\ }\bibfield  {title} {\bibinfo {title} {Defect identification in
  semiconductors with positron annihilation: Experiment and theory},\ }\href
  {https://doi.org/10.1103/RevModPhys.85.1583} {\bibfield  {journal} {\bibinfo
  {journal} {Rev. Mod. Phys.}\ }\textbf {\bibinfo {volume} {85}},\ \bibinfo
  {pages} {1583} (\bibinfo {year} {2013})}\BibitemShut {NoStop}%
\bibitem [{\citenamefont {Hugenschmidt}(2016)}]{Hugenschmidt2016}%
  \BibitemOpen
  \bibfield  {author} {\bibinfo {author} {\bibfnamefont {C.}~\bibnamefont
  {Hugenschmidt}},\ }\href@noop {} {\bibfield  {journal} {\bibinfo  {journal}
  {Surf. Sci. Rep.}\ }\textbf {\bibinfo {volume} {71}},\ \bibinfo {pages} {547}
  (\bibinfo {year} {2016})}\BibitemShut {NoStop}%
\bibitem [{\citenamefont {M.~Leventhal}\ and\ \citenamefont
  {Stang}(1978)}]{leventhal_galacticpositrons}%
  \BibitemOpen
  \bibfield  {author} {\bibinfo {author} {\bibfnamefont {C.~J.~M.}\
  \bibnamefont {M.~Leventhal}}\ and\ \bibinfo {author} {\bibfnamefont {P.~D.}\
  \bibnamefont {Stang}},\ }\bibfield  {title} {\bibinfo {title} {Detection of
  511\,kev positron annihilation radiation from the galactic centre
  direction},\ }\href@noop {} {\bibfield  {journal} {\bibinfo  {journal}
  {Astrophys. J.}\ }\textbf {\bibinfo {volume} {225}},\ \bibinfo {pages} {L11}
  (\bibinfo {year} {1978})}\BibitemShut {NoStop}%
\bibitem [{\citenamefont {Prantzos}\ \emph {et~al.}(2011)\citenamefont
  {Prantzos}, \citenamefont {Boehm}, \citenamefont {Bykov}, \citenamefont
  {Diehl}, \citenamefont {Ferri\`ere}, \citenamefont {Guessoum}, \citenamefont
  {Jean}, \citenamefont {Knoedlseder}, \citenamefont {Marcowith}, \citenamefont
  {Moskalenko}, \citenamefont {Strong},\ and\ \citenamefont
  {Weidenspointner}}]{RevModPhys.83.1001}%
  \BibitemOpen
  \bibfield  {author} {\bibinfo {author} {\bibfnamefont {N.}~\bibnamefont
  {Prantzos}}, \bibinfo {author} {\bibfnamefont {C.}~\bibnamefont {Boehm}},
  \bibinfo {author} {\bibfnamefont {A.}~\bibnamefont {Bykov}}, \bibinfo
  {author} {\bibfnamefont {R.}~\bibnamefont {Diehl}}, \bibinfo {author}
  {\bibfnamefont {K.}~\bibnamefont {Ferri\`ere}}, \bibinfo {author}
  {\bibfnamefont {N.}~\bibnamefont {Guessoum}}, \bibinfo {author}
  {\bibfnamefont {P.}~\bibnamefont {Jean}}, \bibinfo {author} {\bibfnamefont
  {J.}~\bibnamefont {Knoedlseder}}, \bibinfo {author} {\bibfnamefont
  {A.}~\bibnamefont {Marcowith}}, \bibinfo {author} {\bibfnamefont
  {I.}~\bibnamefont {Moskalenko}}, \bibinfo {author} {\bibfnamefont
  {A.}~\bibnamefont {Strong}},\ and\ \bibinfo {author} {\bibfnamefont
  {G.}~\bibnamefont {Weidenspointner}},\ }\bibfield  {title} {\bibinfo {title}
  {The 511 kev emission from positron annihilation in the galaxy},\ }\href
  {https://doi.org/10.1103/RevModPhys.83.1001} {\bibfield  {journal} {\bibinfo
  {journal} {Rev. Mod. Phys.}\ }\textbf {\bibinfo {volume} {83}},\ \bibinfo
  {pages} {1001} (\bibinfo {year} {2011})}\BibitemShut {NoStop}%
\bibitem [{\citenamefont {Wahal}(2008)}]{PETbook}%
  \BibitemOpen
  \bibfield  {author} {\bibinfo {author} {\bibfnamefont {R.~L.}\ \bibnamefont
  {Wahal}},\ }\href@noop {} {\emph {\bibinfo {title} {Principles and Practice
  of Positron Emission Tomography}}}\ (\bibinfo  {publisher} {Lippincott,
  Williams and Wilkins, Philadelphia},\ \bibinfo {year} {2008})\BibitemShut
  {NoStop}%
\bibitem [{\citenamefont {Vaquero}\ and\ \citenamefont
  {Kinahan}(2015)}]{PETnew}%
  \BibitemOpen
  \bibfield  {author} {\bibinfo {author} {\bibfnamefont {J.~J.}\ \bibnamefont
  {Vaquero}}\ and\ \bibinfo {author} {\bibfnamefont {P.}~\bibnamefont
  {Kinahan}},\ }\bibfield  {title} {\bibinfo {title} {Positron emission
  tomography: Current challenges and opportunities for technological advances
  in clinical and preclinical imaging systems},\ }\href
  {https://doi.org/10.1146/annurev-bioeng-071114-040723} {\bibfield  {journal}
  {\bibinfo  {journal} {Ann. Rev. Biomed. Eng.}\ }\textbf {\bibinfo {volume}
  {17}},\ \bibinfo {pages} {385} (\bibinfo {year} {2015})}\BibitemShut
  {NoStop}%
\bibitem [{\citenamefont {\emph{et al.}}(2021)}]{JPET}%
  \BibitemOpen
  \bibfield  {author} {\bibinfo {author} {\bibfnamefont {P.~Moskal.}\ \bibnamefont
  {\emph{et al.}}},\ }\bibfield  {title} {\bibinfo {title} {Positronium imaging
  with the novel multiphoton pet scanner},\ }\href
  {https://doi.org/10.1126/sciadv.abh4394} {\bibfield  {journal} {\bibinfo
  {journal} {Science Advances}\ }\textbf {\bibinfo {volume} {7}},\ \bibinfo
  {pages} {4394} (\bibinfo {year} {2021})}\BibitemShut {NoStop}%
\bibitem [{\citenamefont {Bass}\ \emph {et~al.}(2023)\citenamefont {Bass},
  \citenamefont {Mariazzi}, \citenamefont {Moskal},\ and\ \citenamefont
  {Stepien}}]{RevModPhys.95.021002}%
  \BibitemOpen
  \bibfield  {author} {\bibinfo {author} {\bibfnamefont {S.~D.}\ \bibnamefont
  {Bass}}, \bibinfo {author} {\bibfnamefont {S.}~\bibnamefont {Mariazzi}},
  \bibinfo {author} {\bibfnamefont {P.}~\bibnamefont {Moskal}},\ and\ \bibinfo
  {author} {\bibfnamefont {E.}~\bibnamefont {Stepien}},\ }\bibfield  {title}
  {\bibinfo {title} {Colloquium: Positronium physics and biomedical
  applications},\ }\href {https://doi.org/10.1103/RevModPhys.95.021002}
  {\bibfield  {journal} {\bibinfo  {journal} {Rev. Mod. Phys.}\ }\textbf
  {\bibinfo {volume} {95}},\ \bibinfo {pages} {021002} (\bibinfo {year}
  {2023})}\BibitemShut {NoStop}%
\bibitem [{\citenamefont {Green}(2017{\natexlab{a}})}]{DGG_cool}%
  \BibitemOpen
  \bibfield  {author} {\bibinfo {author} {\bibfnamefont {D.~G.}\ \bibnamefont
  {Green}},\ }\bibfield  {title} {\bibinfo {title} {Positron cooling and
  annihilation in noble gases},\ }\href
  {https://doi.org/10.1103/PhysRevLett.119.203403} {\bibfield  {journal}
  {\bibinfo  {journal} {Phys. Rev. Lett.}\ }\textbf {\bibinfo {volume} {119}},\
  \bibinfo {pages} {203403} (\bibinfo {year} {2017}{\natexlab{a}})}\BibitemShut
  {NoStop}%
\bibitem [{\citenamefont {Green}(2017{\natexlab{b}})}]{DGG_gamcool}%
  \BibitemOpen
  \bibfield  {author} {\bibinfo {author} {\bibfnamefont {D.~G.}\ \bibnamefont
  {Green}},\ }\bibfield  {title} {\bibinfo {title} {Probing positron cooling in
  noble gases via annihilation $\ensuremath{\gamma}$ spectra},\ }\href
  {https://doi.org/10.1103/PhysRevLett.119.203404} {\bibfield  {journal}
  {\bibinfo  {journal} {Phys. Rev. Lett.}\ }\textbf {\bibinfo {volume} {119}},\
  \bibinfo {pages} {203404} (\bibinfo {year} {2017}{\natexlab{b}})}\BibitemShut
  {NoStop}%
\bibitem [{\citenamefont {Ackermann}\ \emph {et~al.}(2016)\citenamefont
  {Ackermann}, \citenamefont {L{\"o}we}, \citenamefont {Dickmann},
  \citenamefont {Mitteneder}, \citenamefont {Sperr}, \citenamefont {Egger},
  \citenamefont {Reiner},\ and\ \citenamefont {Dollinger}}]{AMOC:2016}%
  \BibitemOpen
  \bibfield  {author} {\bibinfo {author} {\bibfnamefont {U.}~\bibnamefont
  {Ackermann}}, \bibinfo {author} {\bibfnamefont {B.}~\bibnamefont {L{\"o}we}},
  \bibinfo {author} {\bibfnamefont {M.}~\bibnamefont {Dickmann}}, \bibinfo
  {author} {\bibfnamefont {J.}~\bibnamefont {Mitteneder}}, \bibinfo {author}
  {\bibfnamefont {P.}~\bibnamefont {Sperr}}, \bibinfo {author} {\bibfnamefont
  {W.}~\bibnamefont {Egger}}, \bibinfo {author} {\bibfnamefont
  {M.}~\bibnamefont {Reiner}},\ and\ \bibinfo {author} {\bibfnamefont
  {G.}~\bibnamefont {Dollinger}},\ }\bibfield  {title} {\bibinfo {title}
  {Four-dimensional positron age-momentum correlation},\ }\href
  {https://doi.org/10.1088/1367-2630/18/11/113030} {\bibfield  {journal}
  {\bibinfo  {journal} {New J. Phys.}\ }\textbf {\bibinfo {volume} {18}},\
  \bibinfo {pages} {113030} (\bibinfo {year} {2016})}\BibitemShut {NoStop}%
\bibitem [{\citenamefont {Siegle}\ \emph {et~al.}(1997)\citenamefont {Siegle},
  \citenamefont {Stoll}, \citenamefont {Castellaz}, \citenamefont {Major},
  \citenamefont {Schneider},\ and\ \citenamefont {Seeger}}]{AMOC:1997}%
  \BibitemOpen
  \bibfield  {author} {\bibinfo {author} {\bibfnamefont {A.}~\bibnamefont
  {Siegle}}, \bibinfo {author} {\bibfnamefont {H.}~\bibnamefont {Stoll}},
  \bibinfo {author} {\bibfnamefont {P.}~\bibnamefont {Castellaz}}, \bibinfo
  {author} {\bibfnamefont {J.}~\bibnamefont {Major}}, \bibinfo {author}
  {\bibfnamefont {H.}~\bibnamefont {Schneider}},\ and\ \bibinfo {author}
  {\bibfnamefont {A.}~\bibnamefont {Seeger}},\ }\bibfield  {title} {\bibinfo
  {title} {Two-dimensional analysis of positron age-momentum correlation (amoc)
  data},\ }\href
  {https://doi.org/http://dx.doi.org/10.1016/S0169-4332(96)01043-4} {\bibfield
  {journal} {\bibinfo  {journal} {App. Surf. Sci.}\ }\textbf {\bibinfo {volume}
  {116}},\ \bibinfo {pages} {140 } (\bibinfo {year} {1997})}\BibitemShut
  {NoStop}%
\bibitem [{\citenamefont {Surko}\ \emph {et~al.}(2005)\citenamefont {Surko},
  \citenamefont {Gribakin},\ and\ \citenamefont {Buckman}}]{Surko:2005}%
  \BibitemOpen
  \bibfield  {author} {\bibinfo {author} {\bibfnamefont {C.~M.}\ \bibnamefont
  {Surko}}, \bibinfo {author} {\bibfnamefont {G.~F.}\ \bibnamefont
  {Gribakin}},\ and\ \bibinfo {author} {\bibfnamefont {S.~J.}\ \bibnamefont
  {Buckman}},\ }\bibfield  {title} {\bibinfo {title} {Low-energy positron
  interactions with atoms and molecules},\ }\href
  {https://doi.org/10.1088/0953-4075/38/6/R01} {\bibfield  {journal} {\bibinfo
  {journal} {J. Phys. B}\ }\textbf {\bibinfo {volume} {38}},\ \bibinfo {pages}
  {R57} (\bibinfo {year} {2005})}\BibitemShut {NoStop}%
\bibitem [{\citenamefont {Gribakin}\ \emph {et~al.}(2010)\citenamefont
  {Gribakin}, \citenamefont {Young},\ and\ \citenamefont
  {Surko}}]{RevModPhys.82.2557}%
  \BibitemOpen
  \bibfield  {author} {\bibinfo {author} {\bibfnamefont {G.~F.}\ \bibnamefont
  {Gribakin}}, \bibinfo {author} {\bibfnamefont {J.~A.}\ \bibnamefont
  {Young}},\ and\ \bibinfo {author} {\bibfnamefont {C.~M.}\ \bibnamefont
  {Surko}},\ }\bibfield  {title} {\bibinfo {title} {Positron-molecule
  interactions: Resonant attachment, annihilation, and bound states},\ }\href
  {https://doi.org/10.1103/RevModPhys.82.2557} {\bibfield  {journal} {\bibinfo
  {journal} {Rev. Mod. Phys.}\ }\textbf {\bibinfo {volume} {82}},\ \bibinfo
  {pages} {2557} (\bibinfo {year} {2010})}\BibitemShut {NoStop}%
\bibitem [{\citenamefont {Green}\ \emph {et~al.}(2014)\citenamefont {Green},
  \citenamefont {Ludlow},\ and\ \citenamefont {Gribakin}}]{Green2014}%
  \BibitemOpen
  \bibfield  {author} {\bibinfo {author} {\bibfnamefont {D.~G.}\ \bibnamefont
  {Green}}, \bibinfo {author} {\bibfnamefont {J.~A.}\ \bibnamefont {Ludlow}},\
  and\ \bibinfo {author} {\bibfnamefont {G.~F.}\ \bibnamefont {Gribakin}},\
  }\bibfield  {title} {\bibinfo {title} {Positron scattering and annihilation
  on noble-gas atoms},\ }\href {https://doi.org/10.1103/PhysRevA.90.032712}
  {\bibfield  {journal} {\bibinfo  {journal} {Phys. Rev. A}\ }\textbf {\bibinfo
  {volume} {90}},\ \bibinfo {pages} {032712} (\bibinfo {year}
  {2014})}\BibitemShut {NoStop}%
\bibitem [{\citenamefont {Hofierka}\ \emph {et~al.}(2022)\citenamefont
  {Hofierka}, \citenamefont {Cunningham}, \citenamefont {Rawlins},
  \citenamefont {Patterson},\ and\ \citenamefont {Green}}]{Hofierka2022}%
  \BibitemOpen
  \bibfield  {author} {\bibinfo {author} {\bibfnamefont {J.}~\bibnamefont
  {Hofierka}}, \bibinfo {author} {\bibfnamefont {B.}~\bibnamefont
  {Cunningham}}, \bibinfo {author} {\bibfnamefont {C.~M.}\ \bibnamefont
  {Rawlins}}, \bibinfo {author} {\bibfnamefont {C.~H.}\ \bibnamefont
  {Patterson}},\ and\ \bibinfo {author} {\bibfnamefont {D.~G.}\ \bibnamefont
  {Green}},\ }\bibfield  {title} {\bibinfo {title} {Many-body theory of
  positron binding to polyatomic molecules},\ }\href@noop {} {\bibfield
  {journal} {\bibinfo  {journal} {Nature}\ }\textbf {\bibinfo {volume} {606}},\
  \bibinfo {pages} {688} (\bibinfo {year} {2022})}\BibitemShut {NoStop}%
\bibitem [{\citenamefont {Iwata}\ \emph {et~al.}(1995)\citenamefont {Iwata},
  \citenamefont {Greaves}, \citenamefont {Murphy}, \citenamefont {Tinkle},\
  and\ \citenamefont {Surko}}]{Iwata1995}%
  \BibitemOpen
  \bibfield  {author} {\bibinfo {author} {\bibfnamefont {K.}~\bibnamefont
  {Iwata}}, \bibinfo {author} {\bibfnamefont {R.~G.}\ \bibnamefont {Greaves}},
  \bibinfo {author} {\bibfnamefont {T.~J.}\ \bibnamefont {Murphy}}, \bibinfo
  {author} {\bibfnamefont {M.~D.}\ \bibnamefont {Tinkle}},\ and\ \bibinfo
  {author} {\bibfnamefont {C.~M.}\ \bibnamefont {Surko}},\ }\bibfield  {title}
  {\bibinfo {title} {Measurements of positron-annihilation rates on
  molecules},\ }\href {https://doi.org/10.1103/PhysRevA.51.473} {\bibfield
  {journal} {\bibinfo  {journal} {Phys. Rev. A}\ }\textbf {\bibinfo {volume}
  {51}},\ \bibinfo {pages} {473} (\bibinfo {year} {1995})}\BibitemShut
  {NoStop}%
\bibitem [{\citenamefont {Dunlop}\ and\ \citenamefont
  {Gribakin}(2006)}]{Dunlop2006}%
  \BibitemOpen
  \bibfield  {author} {\bibinfo {author} {\bibfnamefont {L.~J.~M.}\
  \bibnamefont {Dunlop}}\ and\ \bibinfo {author} {\bibfnamefont {G.~F.}\
  \bibnamefont {Gribakin}},\ }\bibfield  {title} {\bibinfo {title} {Many-body
  theory of gamma spectra from positron–atom annihilation},\ }\href
  {https://doi.org/10.1088/0953-4075/39/7/008} {\bibfield  {journal} {\bibinfo
  {journal} {Journal of Physics B: Atomic, Molecular and Optical Physics}\
  }\textbf {\bibinfo {volume} {39}},\ \bibinfo {pages} {1647} (\bibinfo {year}
  {2006})}\BibitemShut {NoStop}%
\bibitem [{\citenamefont {Green}\ and\ \citenamefont
  {Gribakin}(2013)}]{Green2013}%
  \BibitemOpen
  \bibfield  {author} {\bibinfo {author} {\bibfnamefont {D.~G.}\ \bibnamefont
  {Green}}\ and\ \bibinfo {author} {\bibfnamefont {G.~F.}\ \bibnamefont
  {Gribakin}},\ }\bibfield  {title} {\bibinfo {title} {Positron scattering and
  annihilation in hydrogenlike ions},\ }\href
  {https://doi.org/10.1103/PhysRevA.88.032708} {\bibfield  {journal} {\bibinfo
  {journal} {Phys. Rev. A}\ }\textbf {\bibinfo {volume} {88}},\ \bibinfo
  {pages} {032708} (\bibinfo {year} {2013})}\BibitemShut {NoStop}%
\bibitem [{\citenamefont {Green}\ \emph
  {et~al.}(2010{\natexlab{a}})\citenamefont {Green}, \citenamefont {Saha},
  \citenamefont {Wang}, \citenamefont {Gribakin},\ and\ \citenamefont
  {Surko}}]{Green2010}%
  \BibitemOpen
  \bibfield  {author} {\bibinfo {author} {\bibfnamefont {D.~G.}\ \bibnamefont
  {Green}}, \bibinfo {author} {\bibfnamefont {S.}~\bibnamefont {Saha}},
  \bibinfo {author} {\bibfnamefont {F.}~\bibnamefont {Wang}}, \bibinfo {author}
  {\bibfnamefont {G.~F.}\ \bibnamefont {Gribakin}},\ and\ \bibinfo {author}
  {\bibfnamefont {C.~M.}\ \bibnamefont {Surko}},\ }\bibfield  {title} {\bibinfo
  {title} {Calculation of gamma spectra for positron annihilation on
  molecules},\ }\href
  {https://doi.org/https://doi.org/10.4028/www.scientific.net/msf.666.21}
  {\bibfield  {journal} {\bibinfo  {journal} {Materials Science Forum}\
  }\textbf {\bibinfo {volume} {666}},\ \bibinfo {pages} {21–24} (\bibinfo
  {year} {2010}{\natexlab{a}})}\BibitemShut {NoStop}%
\bibitem [{\citenamefont {Green}\ \emph
  {et~al.}(2010{\natexlab{b}})\citenamefont {Green}, \citenamefont {Saha},
  \citenamefont {Wang}, \citenamefont {Gribakin},\ and\ \citenamefont
  {Surko}}]{DGG_molgammashort}%
  \BibitemOpen
  \bibfield  {author} {\bibinfo {author} {\bibfnamefont {D.~G.}\ \bibnamefont
  {Green}}, \bibinfo {author} {\bibfnamefont {S.}~\bibnamefont {Saha}},
  \bibinfo {author} {\bibfnamefont {F.}~\bibnamefont {Wang}}, \bibinfo {author}
  {\bibfnamefont {G.~F.}\ \bibnamefont {Gribakin}},\ and\ \bibinfo {author}
  {\bibfnamefont {C.~M.}\ \bibnamefont {Surko}},\ }\bibfield  {title} {\bibinfo
  {title} {Calculation of gamma spectra for positron annihilation on
  molecules},\ }\href {https://doi.org/10.4028/www.scientific.net/MSF.666.21}
  {\bibfield  {journal} {\bibinfo  {journal} {Mat.~Sci.~Forum}\ }\textbf
  {\bibinfo {volume} {666}},\ \bibinfo {pages} {21} (\bibinfo {year}
  {2010}{\natexlab{b}})}\BibitemShut {NoStop}%
\bibitem [{\citenamefont {Green}\ \emph {et~al.}(2012)\citenamefont {Green},
  \citenamefont {Saha}, \citenamefont {Wang}, \citenamefont {Gribakin},\ and\
  \citenamefont {Surko}}]{Green_2012}%
  \BibitemOpen
  \bibfield  {author} {\bibinfo {author} {\bibfnamefont {D.~G.}\ \bibnamefont
  {Green}}, \bibinfo {author} {\bibfnamefont {S.}~\bibnamefont {Saha}},
  \bibinfo {author} {\bibfnamefont {F.}~\bibnamefont {Wang}}, \bibinfo {author}
  {\bibfnamefont {G.~F.}\ \bibnamefont {Gribakin}},\ and\ \bibinfo {author}
  {\bibfnamefont {C.~M.}\ \bibnamefont {Surko}},\ }\bibfield  {title} {\bibinfo
  {title} {Effect of positron–atom interactions on the annihilation gamma
  spectra of molecules},\ }\href
  {https://doi.org/10.1088/1367-2630/14/3/035021} {\bibfield  {journal}
  {\bibinfo  {journal} {New Journal of Physics}\ }\textbf {\bibinfo {volume}
  {14}},\ \bibinfo {pages} {035021} (\bibinfo {year} {2012})}\BibitemShut
  {NoStop}%
\bibitem [{\citenamefont {Iwata}\ \emph {et~al.}(1997)\citenamefont {Iwata},
  \citenamefont {Greaves},\ and\ \citenamefont {Surko}}]{Iwata1997_2}%
  \BibitemOpen
  \bibfield  {author} {\bibinfo {author} {\bibfnamefont {K.}~\bibnamefont
  {Iwata}}, \bibinfo {author} {\bibfnamefont {R.~G.}\ \bibnamefont {Greaves}},\
  and\ \bibinfo {author} {\bibfnamefont {C.~M.}\ \bibnamefont {Surko}},\
  }\bibfield  {title} {\bibinfo {title} {$\gamma{}$-ray spectra from positron
  annihilation on atoms and molecules},\ }\href
  {https://doi.org/10.1103/PhysRevA.55.3586} {\bibfield  {journal} {\bibinfo
  {journal} {Phys. Rev. A}\ }\textbf {\bibinfo {volume} {55}},\ \bibinfo
  {pages} {3586} (\bibinfo {year} {1997})}\BibitemShut {NoStop}%
\bibitem [{\citenamefont {Reeth}\ \emph {et~al.}(1996)\citenamefont {Reeth},
  \citenamefont {Humberston}, \citenamefont {Iwata}, \citenamefont {Greaves},\
  and\ \citenamefont {Surko}}]{VanReeth1996}%
  \BibitemOpen
  \bibfield  {author} {\bibinfo {author} {\bibfnamefont {P.~V.}\ \bibnamefont
  {Reeth}}, \bibinfo {author} {\bibfnamefont {J.~W.}\ \bibnamefont
  {Humberston}}, \bibinfo {author} {\bibfnamefont {K.}~\bibnamefont {Iwata}},
  \bibinfo {author} {\bibfnamefont {R.~G.}\ \bibnamefont {Greaves}},\ and\
  \bibinfo {author} {\bibfnamefont {C.~M.}\ \bibnamefont {Surko}},\ }\bibfield
  {title} {\bibinfo {title} {Annihilation in low-energy positron - helium
  scattering},\ }\href {https://doi.org/10.1088/0953-4075/29/12/004} {\bibfield
   {journal} {\bibinfo  {journal} {Journal of Physics B: Atomic, Molecular and
  Optical Physics}\ }\textbf {\bibinfo {volume} {29}},\ \bibinfo {pages} {L465}
  (\bibinfo {year} {1996})}\BibitemShut {NoStop}%
\bibitem [{\citenamefont {Ikabata}\ \emph {et~al.}(2018)\citenamefont
  {Ikabata}, \citenamefont {Aiba}, \citenamefont {Iwanade}, \citenamefont
  {Nishizawa}, \citenamefont {Wang},\ and\ \citenamefont
  {Nakai}}]{Ikabata2018}%
  \BibitemOpen
  \bibfield  {author} {\bibinfo {author} {\bibfnamefont {Y.}~\bibnamefont
  {Ikabata}}, \bibinfo {author} {\bibfnamefont {R.}~\bibnamefont {Aiba}},
  \bibinfo {author} {\bibfnamefont {T.}~\bibnamefont {Iwanade}}, \bibinfo
  {author} {\bibfnamefont {H.}~\bibnamefont {Nishizawa}}, \bibinfo {author}
  {\bibfnamefont {F.}~\bibnamefont {Wang}},\ and\ \bibinfo {author}
  {\bibfnamefont {H.}~\bibnamefont {Nakai}},\ }\bibfield  {title} {\bibinfo
  {title} {{Quantum chemical approach for positron annihilation spectra of
  atoms and molecules beyond plane-wave approximation}},\ }\href
  {https://doi.org/10.1063/1.5019805} {\bibfield  {journal} {\bibinfo
  {journal} {J. Chem. Phys.}\ }\textbf {\bibinfo {volume} {148}},\ \bibinfo
  {pages} {184110} (\bibinfo {year} {2018})}\BibitemShut {NoStop}%
\bibitem [{\citenamefont {Ma}\ \emph {et~al.}(2016)\citenamefont {Ma},
  \citenamefont {Wang}, \citenamefont {Zhu}, \citenamefont {Liu}, \citenamefont
  {Yang},\ and\ \citenamefont {Wang}}]{Ma2016}%
  \BibitemOpen
  \bibfield  {author} {\bibinfo {author} {\bibfnamefont {X.}~\bibnamefont
  {Ma}}, \bibinfo {author} {\bibfnamefont {M.}~\bibnamefont {Wang}}, \bibinfo
  {author} {\bibfnamefont {Y.}~\bibnamefont {Zhu}}, \bibinfo {author}
  {\bibfnamefont {Y.}~\bibnamefont {Liu}}, \bibinfo {author} {\bibfnamefont
  {C.}~\bibnamefont {Yang}},\ and\ \bibinfo {author} {\bibfnamefont
  {D.}~\bibnamefont {Wang}},\ }\bibfield  {title} {\bibinfo {title} {Gamma-ray
  spectra from low-energy positron annihilation processes in molecules},\
  }\href {https://doi.org/10.1103/PhysRevA.94.052709} {\bibfield  {journal}
  {\bibinfo  {journal} {Phys. Rev. A}\ }\textbf {\bibinfo {volume} {94}},\
  \bibinfo {pages} {052709} (\bibinfo {year} {2016})}\BibitemShut {NoStop}%
\bibitem [{\citenamefont {Green}\ and\ \citenamefont
  {Gribakin}(2017)}]{Green2017_comment}%
  \BibitemOpen
  \bibfield  {author} {\bibinfo {author} {\bibfnamefont {D.~G.}\ \bibnamefont
  {Green}}\ and\ \bibinfo {author} {\bibfnamefont {G.~F.}\ \bibnamefont
  {Gribakin}},\ }\bibfield  {title} {\bibinfo {title} {Comment on ``gamma-ray
  spectra from low-energy positron annihilation processes in molecules''},\
  }\href {https://doi.org/10.1103/PhysRevA.95.036701} {\bibfield  {journal}
  {\bibinfo  {journal} {Phys. Rev. A}\ }\textbf {\bibinfo {volume} {95}},\
  \bibinfo {pages} {036701} (\bibinfo {year} {2017})}\BibitemShut {NoStop}%
\bibitem [{\citenamefont {Patterson}(2010)}]{Patterson2010}%
  \BibitemOpen
  \bibfield  {author} {\bibinfo {author} {\bibfnamefont {C.~H.}\ \bibnamefont
  {Patterson}},\ }\bibfield  {title} {\bibinfo {title} {Exciton: a code for
  excitations in materials},\ }\href
  {https://doi.org/10.1080/00268976.2010.505587} {\bibfield  {journal}
  {\bibinfo  {journal} {Molecular Physics}\ }\textbf {\bibinfo {volume}
  {108}},\ \bibinfo {pages} {3181} (\bibinfo {year} {2010})}\BibitemShut
  {NoStop}%
\bibitem [{\citenamefont {Patterson}(2019)}]{Patterson2019}%
  \BibitemOpen
  \bibfield  {author} {\bibinfo {author} {\bibfnamefont {C.~H.}\ \bibnamefont
  {Patterson}},\ }\bibfield  {title} {\bibinfo {title} {Photoabsorption spectra
  of small {Na} clusters: {TDHF} and {BSE} versus {CI} and experiment},\
  }\href@noop {} {\bibfield  {journal} {\bibinfo  {journal} {Phys. Rev.
  Mater.}\ }\textbf {\bibinfo {volume} {3}},\ \bibinfo {pages} {043804}
  (\bibinfo {year} {2019})}\BibitemShut {NoStop}%
\bibitem [{\citenamefont {Patterson}(2020)}]{Patterson2020}%
  \BibitemOpen
  \bibfield  {author} {\bibinfo {author} {\bibfnamefont {C.~H.}\ \bibnamefont
  {Patterson}},\ }\bibfield  {title} {\bibinfo {title} {Density fitting in
  periodic systems: Application to tdhf in diamond and oxides},\ }\href
  {https://doi.org/10.1063/5.0014106} {\bibfield  {journal} {\bibinfo
  {journal} {J. Chem. Phys.}\ }\textbf {\bibinfo {volume} {153}},\ \bibinfo
  {pages} {064107} (\bibinfo {year} {2020})}\BibitemShut {NoStop}%
\bibitem [{\citenamefont {Arthur-Baidoo}\ \emph {et~al.}(2024)\citenamefont
  {Arthur-Baidoo}, \citenamefont {Danielson}, \citenamefont {Surko},
  \citenamefont {Cassidy}, \citenamefont {Gregg}, \citenamefont {Hofierka},
  \citenamefont {Cunningham}, \citenamefont {Patterson},\ and\ \citenamefont
  {Green}}]{ArthurBaidoo2024}%
  \BibitemOpen
  \bibfield  {author} {\bibinfo {author} {\bibfnamefont {E.}~\bibnamefont
  {Arthur-Baidoo}}, \bibinfo {author} {\bibfnamefont {J.~R.}\ \bibnamefont
  {Danielson}}, \bibinfo {author} {\bibfnamefont {C.~M.}\ \bibnamefont
  {Surko}}, \bibinfo {author} {\bibfnamefont {J.~P.}\ \bibnamefont {Cassidy}},
  \bibinfo {author} {\bibfnamefont {S.~K.}\ \bibnamefont {Gregg}}, \bibinfo
  {author} {\bibfnamefont {J.}~\bibnamefont {Hofierka}}, \bibinfo {author}
  {\bibfnamefont {B.}~\bibnamefont {Cunningham}}, \bibinfo {author}
  {\bibfnamefont {C.~H.}\ \bibnamefont {Patterson}},\ and\ \bibinfo {author}
  {\bibfnamefont {D.~G.}\ \bibnamefont {Green}},\ }\bibfield  {title} {\bibinfo
  {title} {Positron annihilation and binding in aromatic and other ring
  molecules},\ }\href {https://doi.org/10.1103/PhysRevA.109.062801} {\bibfield
  {journal} {\bibinfo  {journal} {Phys. Rev. A}\ }\textbf {\bibinfo {volume}
  {109}},\ \bibinfo {pages} {062801} (\bibinfo {year} {2024})}\BibitemShut
  {NoStop}%
\bibitem [{\citenamefont {Hofierka}\ \emph {et~al.}(2024)\citenamefont
  {Hofierka}, \citenamefont {Cunningham},\ and\ \citenamefont
  {Green}}]{Hofierka2024}%
  \BibitemOpen
  \bibfield  {author} {\bibinfo {author} {\bibfnamefont {J.}~\bibnamefont
  {Hofierka}}, \bibinfo {author} {\bibfnamefont {B.}~\bibnamefont
  {Cunningham}},\ and\ \bibinfo {author} {\bibfnamefont {D.~G.}\ \bibnamefont
  {Green}},\ }\bibfield  {title} {\bibinfo {title} {Many-body theory
  calculations of positron binding to hydrogen cyanide},\ }\href
  {https://doi.org/{10.1140/epjd/s10053-024-00810-0}} {\bibfield  {journal}
  {\bibinfo  {journal} {Eur. Phys. J. D}\ }\textbf {\bibinfo {volume} {78}},\
  \bibinfo {pages} {37} (\bibinfo {year} {2024})}\BibitemShut {NoStop}%
\bibitem [{\citenamefont {Cassidy}\ \emph
  {et~al.}(2024{\natexlab{a}})\citenamefont {Cassidy}, \citenamefont
  {Hofierka}, \citenamefont {Cunningham}, \citenamefont {Rawlins},
  \citenamefont {Patterson},\ and\ \citenamefont {Green}}]{Cassidy2023}%
  \BibitemOpen
  \bibfield  {author} {\bibinfo {author} {\bibfnamefont {J.~P.}\ \bibnamefont
  {Cassidy}}, \bibinfo {author} {\bibfnamefont {J.}~\bibnamefont {Hofierka}},
  \bibinfo {author} {\bibfnamefont {B.}~\bibnamefont {Cunningham}}, \bibinfo
  {author} {\bibfnamefont {C.~M.}\ \bibnamefont {Rawlins}}, \bibinfo {author}
  {\bibfnamefont {C.~H.}\ \bibnamefont {Patterson}},\ and\ \bibinfo {author}
  {\bibfnamefont {D.~G.}\ \bibnamefont {Green}},\ }\bibfield  {title} {\bibinfo
  {title} {Many-body theory calculations of positron binding to halogenated
  hydrocarbons},\ }\href {https://doi.org/10.1103/PhysRevA.109.L040801}
  {\bibfield  {journal} {\bibinfo  {journal} {Phys. Rev. A}\ }\textbf {\bibinfo
  {volume} {109}},\ \bibinfo {pages} {L040801} (\bibinfo {year}
  {2024}{\natexlab{a}})}\BibitemShut {NoStop}%
\bibitem [{\citenamefont {Rawlins}\ \emph {et~al.}(2023)\citenamefont
  {Rawlins}, \citenamefont {Hofierka}, \citenamefont {Cunningham},
  \citenamefont {Patterson},\ and\ \citenamefont {Green}}]{Rawlins2023}%
  \BibitemOpen
  \bibfield  {author} {\bibinfo {author} {\bibfnamefont {C.~M.}\ \bibnamefont
  {Rawlins}}, \bibinfo {author} {\bibfnamefont {J.}~\bibnamefont {Hofierka}},
  \bibinfo {author} {\bibfnamefont {B.}~\bibnamefont {Cunningham}}, \bibinfo
  {author} {\bibfnamefont {C.~H.}\ \bibnamefont {Patterson}},\ and\ \bibinfo
  {author} {\bibfnamefont {D.~G.}\ \bibnamefont {Green}},\ }\bibfield  {title}
  {\bibinfo {title} {Many-body theory calculations of positron scattering and
  annihilation in {H$_2$, N$_2$, and CH$_4$}},\ }\href
  {https://doi.org/10.1103/PhysRevLett.130.263001} {\bibfield  {journal}
  {\bibinfo  {journal} {Phys. Rev. Lett.}\ }\textbf {\bibinfo {volume} {130}},\
  \bibinfo {pages} {263001} (\bibinfo {year} {2023})}\BibitemShut {NoStop}%
\bibitem [{\citenamefont {Cassidy}\ \emph
  {et~al.}(2024{\natexlab{b}})\citenamefont {Cassidy}, \citenamefont
  {Hofierka}, \citenamefont {Cunningham},\ and\ \citenamefont
  {Green}}]{Cassidy2024_2}%
  \BibitemOpen
  \bibfield  {author} {\bibinfo {author} {\bibfnamefont {J.~P.}\ \bibnamefont
  {Cassidy}}, \bibinfo {author} {\bibfnamefont {J.}~\bibnamefont {Hofierka}},
  \bibinfo {author} {\bibfnamefont {B.}~\bibnamefont {Cunningham}},\ and\
  \bibinfo {author} {\bibfnamefont {D.~G.}\ \bibnamefont {Green}},\ }\bibfield
  {title} {\bibinfo {title} {Many-body theory calculations of positronic-bonded
  molecular dianions},\ }\href {https://doi.org/10.1063/5.0188719} {\bibfield
  {journal} {\bibinfo  {journal} {J. Chem. Phys.}\ }\textbf {\bibinfo {volume}
  {160}},\ \bibinfo {pages} {084304} (\bibinfo {year}
  {2024}{\natexlab{b}})}\BibitemShut {NoStop}%
\bibitem [{Note1()}]{Note1}%
  \BibitemOpen
  \bibinfo {note} {For atoms the shape of the spectra changes slowly at low
  positron energy \cite {Green2015,DGG:2017:ef}. On the other hand, for
  molecules the positron bound state can be highly anisotropic.}\BibitemShut
  {Stop}%
\bibitem [{\citenamefont {Berestetskii}\ \emph {et~al.}(1982)\citenamefont
  {Berestetskii}, \citenamefont {Lifshitz},\ and\ \citenamefont
  {Pitaevskii}}]{Berestetskii1982}%
  \BibitemOpen
  \bibfield  {author} {\bibinfo {author} {\bibfnamefont {V.}~\bibnamefont
  {Berestetskii}}, \bibinfo {author} {\bibfnamefont {E.}~\bibnamefont
  {Lifshitz}},\ and\ \bibinfo {author} {\bibfnamefont {L.}~\bibnamefont
  {Pitaevskii}},\ }\href@noop {} {\emph {\bibinfo {title} {Quantum
  Electrodynamics}}},\ \bibinfo {edition} {2nd}\ ed.\ (\bibinfo  {publisher}
  {Pergamon, Oxford},\ \bibinfo {year} {1982})\BibitemShut {NoStop}%
\bibitem [{\citenamefont {McMurchie}\ and\ \citenamefont
  {Davidson}(1978)}]{McMurchie1978}%
  \BibitemOpen
  \bibfield  {author} {\bibinfo {author} {\bibfnamefont {L.~E.}\ \bibnamefont
  {McMurchie}}\ and\ \bibinfo {author} {\bibfnamefont {E.~R.}\ \bibnamefont
  {Davidson}},\ }\bibfield  {title} {\bibinfo {title} {One- and two-electron
  integrals over cartesian gaussian functions},\ }\href
  {https://doi.org/{10.1016/0021-9991(78)90092-X}} {\bibfield  {journal}
  {\bibinfo  {journal} {J. Comp. Phys.}\ }\textbf {\bibinfo {volume} {26}},\
  \bibinfo {pages} {218} (\bibinfo {year} {1978})}\BibitemShut {NoStop}%
\bibitem [{\citenamefont {Abramowitz}\ and\ \citenamefont
  {Stegun}(1965)}]{Abramowicz}%
  \BibitemOpen
  \bibfield  {author} {\bibinfo {author} {\bibfnamefont {M.}~\bibnamefont
  {Abramowitz}}\ and\ \bibinfo {author} {\bibfnamefont {I.}~\bibnamefont
  {Stegun}},\ }\href@noop {} {\emph {\bibinfo {title} {Handbook of Mathematical
  Functions: With Formulas, Graphs, and Mathematical Tables}}},\ Applied
  mathematics series\ (\bibinfo  {publisher} {Dover Publications},\ \bibinfo
  {year} {1965})\BibitemShut {NoStop}%
\bibitem [{\citenamefont {Green}\ and\ \citenamefont
  {Gribakin}(2018)}]{DGG:2017:ef}%
  \BibitemOpen
  \bibfield  {author} {\bibinfo {author} {\bibfnamefont {D.~G.}\ \bibnamefont
  {Green}}\ and\ \bibinfo {author} {\bibfnamefont {G.~F.}\ \bibnamefont
  {Gribakin}},\ }\bibfield  {title} {\bibinfo {title} {Enhancement factors for
  positron annihilation on valence and core orbitals of noble-gas atoms},\
  }\href {https://doi.org/"10.1007/978-3-319-74582-4_14"} {\bibfield  {journal}
  {\bibinfo  {journal} {Concepts, Methods and Applications of Quantum Systems
  in Chemistry and Physics, Prog. Theor. Chem. and Phys.}\ }\textbf {\bibinfo
  {volume} {31}},\ \bibinfo {pages} {243} (\bibinfo {year} {2018})}\BibitemShut
  {NoStop}%
\bibitem [{\citenamefont {Green}\ \emph {et~al.}(2018)\citenamefont {Green},
  \citenamefont {Swann},\ and\ \citenamefont {Gribakin}}]{DGG:2018:PRL}%
  \BibitemOpen
  \bibfield  {author} {\bibinfo {author} {\bibfnamefont {D.~G.}\ \bibnamefont
  {Green}}, \bibinfo {author} {\bibfnamefont {A.~R.}\ \bibnamefont {Swann}},\
  and\ \bibinfo {author} {\bibfnamefont {G.~F.}\ \bibnamefont {Gribakin}},\
  }\bibfield  {title} {\bibinfo {title} {Many-body theory for positronium-atom
  interactions},\ }\href {https://doi.org/10.1103/PhysRevLett.120.183402}
  {\bibfield  {journal} {\bibinfo  {journal} {Phys. Rev. Lett.}\ }\textbf
  {\bibinfo {volume} {120}},\ \bibinfo {pages} {183402} (\bibinfo {year}
  {2018})}\BibitemShut {NoStop}%
\bibitem [{\citenamefont {Swann}\ \emph {et~al.}(2023)\citenamefont {Swann},
  \citenamefont {Green},\ and\ \citenamefont {Gribakin}}]{Swann:2023}%
  \BibitemOpen
  \bibfield  {author} {\bibinfo {author} {\bibfnamefont {A.~R.}\ \bibnamefont
  {Swann}}, \bibinfo {author} {\bibfnamefont {D.~G.}\ \bibnamefont {Green}},\
  and\ \bibinfo {author} {\bibfnamefont {G.~F.}\ \bibnamefont {Gribakin}},\
  }\bibfield  {title} {\bibinfo {title} {Many-body theory of positronium
  scattering and pickoff annihilation in noble-gas atoms},\ }\href
  {https://doi.org/10.1103/PhysRevA.107.042802} {\bibfield  {journal} {\bibinfo
   {journal} {Phys. Rev. A}\ }\textbf {\bibinfo {volume} {107}},\ \bibinfo
  {pages} {042802} (\bibinfo {year} {2023})}\BibitemShut {NoStop}%
\bibitem [{\citenamefont {Bergami}\ \emph {et~al.}(2022)\citenamefont
  {Bergami}, \citenamefont {Santana}, \citenamefont {Charry~Martinez},
  \citenamefont {Reyes}, \citenamefont {Coutinho},\ and\ \citenamefont
  {Varella}}]{Bergami2022}%
  \BibitemOpen
  \bibfield  {author} {\bibinfo {author} {\bibfnamefont {M.}~\bibnamefont
  {Bergami}}, \bibinfo {author} {\bibfnamefont {A.~L.~D.}\ \bibnamefont
  {Santana}}, \bibinfo {author} {\bibfnamefont {J.}~\bibnamefont
  {Charry~Martinez}}, \bibinfo {author} {\bibfnamefont {A.}~\bibnamefont
  {Reyes}}, \bibinfo {author} {\bibfnamefont {K.}~\bibnamefont {Coutinho}},\
  and\ \bibinfo {author} {\bibfnamefont {M.~T. d.~N.}\ \bibnamefont
  {Varella}},\ }\bibfield  {title} {\bibinfo {title} {Multicomponent quantum
  mechanics/molecular mechanics study of hydrated positronium},\ }\href
  {https://doi.org/10.1021/acs.jpcb.1c10124} {\bibfield  {journal} {\bibinfo
  {journal} {J. Phys. Chem. B}\ }\textbf {\bibinfo {volume} {126}},\ \bibinfo
  {pages} {2699} (\bibinfo {year} {2022})}\BibitemShut {NoStop}%
\bibitem [{\citenamefont {Cassidy}(2024)}]{Cassidy:thesis}%
  \BibitemOpen
  \bibfield  {author} {\bibinfo {author} {\bibfnamefont {J.~P.}\ \bibnamefont
  {Cassidy}},\ }\href@noop {} {Ph.D. thesis},\ \bibinfo  {school} {Queen's
  University Belfast} (\bibinfo {year} {2024})\BibitemShut {NoStop}%
\bibitem [{\citenamefont {Kendall}\ \emph {et~al.}(1992)\citenamefont
  {Kendall}, \citenamefont {Dunning~Jr},\ and\ \citenamefont
  {Harrison}}]{Dunning1992}%
  \BibitemOpen
  \bibfield  {author} {\bibinfo {author} {\bibfnamefont {R.~A.}\ \bibnamefont
  {Kendall}}, \bibinfo {author} {\bibfnamefont {T.}~\bibnamefont
  {Dunning~Jr}},\ and\ \bibinfo {author} {\bibfnamefont {R.~J.}\ \bibnamefont
  {Harrison}},\ }\bibfield  {title} {\bibinfo {title} {Electron affinities of
  the first-row atoms revisited. systematic basis sets and wave functions},\
  }\href@noop {} {\bibfield  {journal} {\bibinfo  {journal} {J. Chem. Phys.}\
  }\textbf {\bibinfo {volume} {96}},\ \bibinfo {pages} {6796} (\bibinfo {year}
  {1992})}\BibitemShut {NoStop}%
\bibitem [{xmg(2024)}]{xmgrace}%
  \BibitemOpen
  \href@noop {} {}\bibinfo {howpublished}
  {https://plasma-gate.weizmann.ac.il/Grace/} (\bibinfo {year}
  {2024})\BibitemShut {NoStop}%
\end{thebibliography}

\onecolumngrid
\appendix

\section{Calculation of the annihilation amplitude integral}\label{app1}
This appendix contains an outline of the procedure followed to evaluate the annihilation amplitude in the independent particle approximation, $A^{(0)}_{n\varepsilon}(\mathbf{P})$. 
Substituting the Gaussian expansions of the electron and positron wavefunctions from Equations \ref{eq:gaussian_expn} in Equation~\ref{eq:ann_amp} yields the following expression for the annihilation amplitude:
\begin{equation}
A^{(0)}_{n\varepsilon}(\mathbf{P}) = \sum_i \sum_j C_i^{\varepsilon} C^{(n)}_j \prod_{\mu=x,y,z} I^{(\mu)}_{n; i j}(P_\mu) , \label{eq5}
\end{equation}
where total momentum $\mathbf{P}=(P_x,P_y,P_z)$, the product over $\mu = x,y,z$ accounts for the components of the wavefunction in each of the three Cartesian directions, the sum over $i$ ($j$) runs over all positron (electron) basis functions, $C_i^{\varepsilon}$ and $C_j^{(n)}$ are expansion coefficients, and
\begin{align}
I^{(\mu)}_{n; i j}(P_\mu)&=\!\!\int_{-\infty}^\infty \!\!\!(\mu\!-\!\mu_i)^{n^\mu_i}\! (\mu\!-\!\mu_j)^{n^\mu_j} e^{-iP_\mu\mu} e^{-\zeta_i(\mu\!-\!\mu_i)^2}e^{-\zeta_j(\mu\!-\!\mu_j)^2}\!d\mu, \\
&=\!e^{-\lambda_{ij}(\mu_i\!-\!\mu_j)^2}\int_{-\infty}^\infty \!(\mu\!-\!\mu_i)^{n^\mu_i} \!(\mu\!-\!\mu_j)^{n^\mu_j} e^{-iP_\mu\mu}  e^{-(\zeta_i+\zeta_j)(\mu\!-\!\mu_{ij})^2}\!d\mu,
\label{def_int_i}\end{align}
where $n_i^{\mu}$ are angular momentum components, $\zeta_i$ are Gaussian exponents, $\mu_i$ are components of the centres of the Gaussian functions, and the last equation is obtained using the Gaussian product rule with
\begin{equation}
\lambda_{ij}=\frac{\zeta_i\zeta_j}{\zeta_i+\zeta_j},\qquad \mu_{ij}=\frac{\zeta_i\mu_i+\zeta_j\mu_j}{\zeta_i+\zeta_j}.
\end{equation}

To solve the integral in Equation~\ref{def_int_i}, we make a Hermite expansion of the polynomial factors:
\begin{align}\label{eq:hermite_expansion}
(\mu-\mu_i)^{n^\mu_i} (\mu-\mu_j)^{n^\mu_j}   = \sum_{s^\mu_{ij}=0}^{n^\mu_i+n^\mu_j} E_{s^\mu_{ij}}^{n^\mu_i n^\mu_j} \Lambda_{s^\mu_{ij}}(\mu-\mu_{ij};\zeta_i+\zeta_j),
\end{align}
where 
\begin{equation}
    \Lambda_{s^\mu_{ij}}(\mu-\mu_{ij};\zeta_i+\zeta_j)=(\zeta_i+\zeta_j)^{s^\mu_{ij}/2} H_{s^\mu_{ij}}\left[(\zeta_i+\zeta_j)^{1/2}(\mu-\mu_{ij})\right]
\end{equation}
and $H_k$ is the $k^{\rm th}$ Hermite polynomial:
\begin{align}
H_k(z)=(-1)^k e^{z^2} \frac{d^k}{dz^k} e^{-z^2}.
\end{align}
The expansion coefficients $E_{s^\mu_{ij}}^{n^\mu_i n^\mu_j}$  satisfy the McMurchie-Davidson recurrence relations \cite{McMurchie1978}:
\begin{align}
E_{s^\mu_{ij}}^{n^\mu_i+1, n^\mu_j}=\frac{1}{2(\zeta_i+\zeta_j)} E_{s^\mu_{ij}-1}^{n^\mu_i n^\mu_j} + (\mu_{ij}-\mu_i) E_{s^\mu_{ij}}^{n^\mu_i n^\mu_j} + (s^\mu_{ij}+1) E_{s^\mu_{ij}+1}^{n^\mu_i n^\mu_j} , \\
E_{s^\mu_{ij}}^{n^\mu_i, n^\mu_j+1}=\frac{1}{2(\zeta_i+\zeta_j)} E_{s^\mu_{ij}-1}^{n^\mu_i n^\mu_j} + (\mu_{ij}-\mu_j) E_{s^\mu_{ij}}^{n^\mu_i n^\mu_j} + (s^\mu_{ij}+1) E_{s^\mu_{ij}+1}^{n^\mu_i n^\mu_j} , 
\end{align}
with $E^{00}_0=1$.
Substituting Equation~\ref{eq:hermite_expansion} into Equation~\ref{def_int_i} yields
\begin{align}
I^{(\mu)}_{n; i j}(P_\mu)
&= e^{-\lambda_{ij}(\mu_i-\mu_j)^2}\sum_{s^\mu_{ij}=0}^{n^\mu_i+n^\mu_j} E_{s^\mu_{ij}}^{n^\mu_i n^\mu_j} 
\int_{-\infty}^\infty  e^{-iP_\mu\mu}   \Lambda_{s^\mu_{ij}}(\mu-\mu_{ij};\zeta_i+\zeta_j)e^{-(\zeta_i+\zeta_j)(\mu-\mu_{ij})^2} \,d\mu. \label{eq:tfygv}
\end{align}

Now, a further transformation can be made using the identity
\begin{align}
 \Lambda_{s^\mu_{ij}}(\mu-\mu_{ij};\zeta_i+\zeta_j)e^{-(\zeta_i+\zeta_j)(\mu-\mu_{ij})^2}= \left( \frac{\partial}{\partial\mu_{ij}} \right)^{s^\mu_{ij}}e^{-(\zeta_i+\zeta_j)(\mu-\mu_{ij})^2},\label{eq:Lambda_deriv}
\end{align}
so that we can rewrite Equation~\ref{eq:tfygv} as
\begin{align}
I^{(\mu)}_{n; i j}(P_\mu)
&=e^{-\lambda_{ij}(\mu_i-\mu_j)^2} \sum_{s^\mu_{ij}=0}^{n^\mu_i+n^\mu_j} E_{s^\mu_{ij}}^{n^\mu_i n^\mu_j} \left( \frac{\partial}{\partial\mu_{ij}} \right)^{s^\mu_{ij}} 
\int_{-\infty}^\infty  e^{-iP_\mu\mu}  e^{-(\zeta_i+\zeta_j)(\mu-\mu_{ij})^2}\,d\mu .\label{eq:ann_amp_int1}
\end{align}
We now make a slight modification to the integral in Equation \ref{eq:ann_amp_int1} to match the form of a common integral encountered in quantum field theory by taking a constant factor of $\exp(-iP_{\mu}\mu_{ij})$ outside of the integral:
\begin{align}
I^{(\mu)}_{n; i j}(P_\mu)
&=e^{-\lambda_{ij}(\mu_i-\mu_j)^2}\sum_{s^\mu_{ij}=0}^{n^\mu_i+n^\mu_j} E_{s^\mu_{ij}}^{n^\mu_i n^\mu_j} \left( \frac{\partial}{\partial\mu_{ij}} \right)^{s^\mu_{ij}} 
e^{-iP_\mu \mu_{ij}}
\int_{-\infty}^\infty  e^{-iP_\mu(\mu-\mu_{ij})}  e^{-(\zeta_i+\zeta_j)(\mu-\mu_{ij})^2}\,d\mu .
\end{align}
Now the integral can be solved using the identity
\begin{gather}
\int_{-\infty}^\infty  e^{-iP_\mu\xi}  e^{-(\zeta_i+\zeta_j)\xi^2}\,d\xi = \sqrt{\frac{\pi}{\zeta_i+\zeta_j}}\, e^{-P_\mu^2/4(\zeta_i+\zeta_j)},
\end{gather}
and we have
\begin{align}
I^{(\mu)}_{n; i j}(P_\mu) = 
 \sqrt{\frac{\pi}{\zeta_i+\zeta_j}} \,
e^{-\lambda_{ij}(\mu_i-\mu_j)^2}
e^{-P_\mu^2/4(\zeta_i+\zeta_j)}
 \sum_{s^\mu_{ij}=0}^{n^\mu_i+n^\mu_j} E_{s^\mu_{ij}}^{n^\mu_i n^\mu_j} 
\left( \frac{\partial}{\partial\mu_{ij}} \right)^{s^\mu_{ij}} 
e^{-iP_\mu \mu_{ij}} . \label{eq:ctvybunh}
\end{align}

Finally, we substitute Equation~\ref{eq:ctvybunh} into Equation~\ref{eq5} to obtain the expression for the annihilation amplitude:
\begin{equation}
A_{n\varepsilon}^{(0)}(\mathbf P) = \sum_i \sum_j C_i^{\varepsilon} C^{(n)}_j \prod_{\mu=x,y,z} \sqrt{\frac{\pi}{\zeta_i+\zeta_j}} \,
e^{-\lambda_{ij}(\mu_i-\mu_j)^2}
e^{-P_\mu^2/4(\zeta_i+\zeta_j)}
 \sum_{s^\mu_{ij}=0}^{n^\mu_i+n^\mu_j} E_{s^\mu_{ij}}^{n^\mu_i n^\mu_j} 
\left( \frac{\partial}{\partial\mu_{ij}} \right)^{s^\mu_{ij}} 
e^{-iP_\mu \mu_{ij}}. \label{ann_amp}
\end{equation}

\section{Calculating the {$\gamma$} spectrum}\label{app:gammaspectrum}

This appendix outlines the solution of the integral in Equation (\ref{eqn:gamspec}). First, we take the square modulus of Equation~\ref{ann_amp} and integrate over solid angle  $\Omega_{\mathbf P}$:
\begin{align}
\int\lvert A^{(0)}_{n\varepsilon}(\mathbf{P})\rvert^2 \,d\Omega_{\mathbf P}
&= \sum_{i,j,i',j'} C_i^{\varepsilon} C_j^{(n)} C_{i'}^{\varepsilon *} C_{j'}^{(n)*}
\int I^{(x)}_{n;ij}(P_x) I^{(y)}_{n;ij}(P_y) I^{(z)}_{n;ij}(P_z) 
I^{(x)*}_{n;i'j'}(P_x) I^{(y)*}_{n;i'j'}(P_y) I^{(z)*}_{n;i'j'}(P_z)\,d\Omega_{\mathbf P} \notag\\
&= \sum_{i,j,i',j'} C_i^{\varepsilon} C_j^{(n)} C_{i'}^{\varepsilon *} C_{j'}^{(n)*}
\frac{\pi^3}{[(\zeta_i+\zeta_j)(\zeta_{i'}+\zeta_{j'})]^{3/2}} 
e^{-\lambda_{ij}\lvert \mathbf r_i-\mathbf r_j\rvert^2}
e^{-\lambda_{i'j'}\lvert \mathbf r_{i'}-\mathbf r_{j'}\rvert^2} 
\notag\\
&\quad{}\times e^{-P^2/4(\zeta_i+\zeta_j)} e^{-P^2/4(\zeta_{i'}+\zeta_{j'})}\notag\\
&\quad{}\times \sum_{\substack{s^x_{ij},s^y_{ij},s^z_{ij},\\ s^x_{i'j'},s^y_{i'j'},s^z_{i'j'}}}
E_{s^x_{ij}}^{n^x_i n^x_j} 
E_{s^y_{ij}}^{n^y_i n^y_j} 
E_{s^z_{ij}}^{n^z_i n^z_j} 
E_{s^x_{i'j'}}^{n^x_{i'} n^x_{j'}} 
E_{s^y_{i'j'}}^{n^y_{i'} n^y_{j'}} 
E_{s^z_{i'j'}}^{n^z_{i'} n^z_{j'}} \notag\\
&\quad{}\times 
\left( \frac{\partial}{\partial x_{ij}} \right)^{s^x_{ij}} 
\left( \frac{\partial}{\partial y_{ij}} \right)^{s^y_{ij}} 
\left( \frac{\partial}{\partial z_{ij}} \right)^{s^z_{ij}} 
\left( \frac{\partial}{\partial x_{i'j'}} \right)^{s^x_{i'j'}} 
\left( \frac{\partial}{\partial y_{i'j'}} \right)^{s^y_{i'j'}} 
\left( \frac{\partial}{\partial z_{i'j'}} \right)^{s^z_{i'j'}} \notag\\
&\quad{}\times 
\int \exp(-i \mathbf P \cdot \mathbf R_{iji'j'})\,d\Omega_{\mathbf P}, \label{eq:hsfhnfbg}
\end{align}
where $P^2=P_x^2+P_y^2+P_z^2$ and $\mathbf R_{iji'j'}=\mathbf r_{ij}-\mathbf r_{i'j'}=(x_{ij}-x_{i'j'},y_{ij}-y_{i'j'},z_{ij}-z_{i'j'})$.
For $\mathbf R_{iji'j'}=\boldsymbol 0$, the integral at the end of Equation~\ref{eq:hsfhnfbg} is
\begin{align}
\int \exp(-i \mathbf P \cdot \mathbf R_{iji'j'})\,d\Omega_{\mathbf P} = \int d\Omega_{\mathbf P}=4\pi.
\end{align}

For $\mathbf R_{iji'j'}\neq\boldsymbol 0$ the integration is carried out analytically in spherical polar coordinates. The integrand depends only on the angle between $\mathbf{P}$ and $\mathbf{R}_{iji'j'}$, so we can choose to align the polar axis with $\mathbf R_{iji'j'}$:
\begin{align}
\int \exp (-i\mathbf P \cdot \mathbf R_{iji'j'})\,d\Omega_{\mathbf P} 
&= \int_0^{2\pi}\!\! \int_0^\pi \exp(-iPR_{iji'j'}\cos\theta_{\mathbf P}) \sin\theta_{\mathbf P} \, d\theta_{\mathbf P} \, d\phi_{\mathbf P}, \notag\\
&=2\pi \int_0^\pi \exp(-iPR_{iji'j'}\cos\theta_{\mathbf P}) \sin\theta_{\mathbf P} \, d\theta_{\mathbf P},\notag
\\&= \frac{4\pi}{PR_{iji'j'}} \sin PR_{iji'j'} ,
\end{align}
where $R_{iji'j'}=\lvert \mathbf R_{iji'j'} \rvert$.
Using the fact that the zeroth spherical Bessel function $j_0$ is given by $j_0(0)=1$, $j_0(z)=z^{-1}\sin z$ ($\lvert z \rvert>0$), we can write in general that
\begin{align}
\int \exp (-i\mathbf P \cdot \mathbf R_{iji'j'})\,d\Omega_{\mathbf P} = 4\pi j_0(PR_{iji'j'}). \label{eq:ctvybu}
\end{align}

Having found an expression for this integral, we now turn our attention to the partial
derivatives in Eqn.~\ref{eq:hsfhnfbg}.
By definition of $\mathbf R_{iji'j'}$, we have
\begin{align}
\frac{\partial}{\partial x_{ij}} &\equiv \frac{\partial}{\partial { X}_{iji'j'}} , 
&\frac{\partial}{\partial y_{ij}} &\equiv \frac{\partial}{\partial { Y}_{iji'j'}} , 
&\frac{\partial}{\partial z_{ij}} &\equiv \frac{\partial}{\partial { Z}_{iji'j'}} , \notag\\
\frac{\partial}{\partial x_{i'j'}} &\equiv -\frac{\partial}{\partial { X}_{iji'j'}}, 
&\frac{\partial}{\partial y_{i'j'}} &\equiv -\frac{\partial}{\partial { Y}_{iji'j'}},  
&\frac{\partial}{\partial z_{i'j'}} &\equiv -\frac{\partial}{\partial { Z}_{iji'j'}}, \label{eq:ctvygb}
\end{align}
where $X_{iji'j'}=x_{ij}-x_{i'j'}$, $Y_{iji'j'}=y_{ij}-y_{i'j'}$ and $Z_{iji'j'}=z_{ij}-z_{i'j'}$.

Substituting Eqns.~\ref{eq:ctvybu} and \ref{eq:ctvygb} into Eqn.~\ref{eq:hsfhnfbg}, we find
\begin{align}
\int \lvert A^{(0)}_{n\varepsilon}(\mathbf{P})\rvert^2 \,d\Omega_{\mathbf P}
&= \sum_{i,j,i',j'} C_i^{\varepsilon} C_j^{(n)} C_{i'}^{\varepsilon *} C_{j'}^{(n)*}
\frac{4\pi^4}{[(\zeta_i+\zeta_j)(\zeta_{i'}+\zeta_{j'})]^{3/2}} 
e^{-\lambda_{ij}\lvert \mathbf r_i-\mathbf r_j\rvert^2}
e^{-\lambda_{i'j'}\lvert \mathbf r_{i'}-\mathbf r_{j'}\rvert^2} 
\notag\\
&\quad{}\times e^{-P^2/4(\zeta_i+\zeta_j)} e^{-P^2/4(\zeta_{i'}+\zeta_{j'})}\notag\\
&\quad{}\times \sum_{\substack{s^x_{ij},s^y_{ij},s^z_{ij},\\ s^x_{i'j'},s^y_{i'j'},s^z_{i'j'}}}
E_{s^x_{ij}}^{n^x_i n^x_j} 
E_{s^y_{ij}}^{n^y_i n^y_j} 
E_{s^z_{ij}}^{n^z_i n^z_j} 
E_{s^x_{i'j'}}^{n^x_{i'} n^x_{j'}} 
E_{s^y_{i'j'}}^{n^y_{i'} n^y_{j'}} 
E_{s^z_{i'j'}}^{n^z_{i'} n^z_{j'}}
(-1)^{s^x_{i'j'}+s^y_{i'j'}+s^z_{i'j'}} \notag\\
&\quad{}\times 
\left( \frac{\partial}{\partial X_{iji'j'}} \right)^{s^x_{ij}+s^x_{i'j'}} 
\left( \frac{\partial}{\partial Y_{iji'j'}} \right)^{s^y_{ij}+s^y_{i'j'}} 
\left( \frac{\partial}{\partial Z_{iji'j'}} \right)^{s^z_{ij}+s^z_{i'j'}} 
j_0(PR_{iji'j'}). \label{eq;nbfihufvij}
\end{align}
To evaluate the partial derivatives, we define the following quantity:
\begin{align}
\mathfrak{j}_n(P,R_{iji'j'}) = \left( \frac{P}{R_{iji'j'}}\right)^n j_n(PR_{iji'j'}),\label{eq:jfrak_def}
\end{align}
noting that for zeroth order, we have $\mathfrak j_0(P,R_{iji'j'})=j_0(PR_{iji'j'})$. 
Equation~\ref{eq;nbfihufvij} becomes
\begin{align} \label{eq:gjffjjnf}
\int \lvert A^{(0)}_{n\varepsilon}(\mathbf{P})\rvert^2 \,d\Omega_{\mathbf P}
&= \sum_{i,j,i',j'} C_i^{\varepsilon} C_j^{(n)} C_{i'}^{\varepsilon *} C_{j'}^{(n)*}
\frac{4\pi^4}{[(\zeta_i+\zeta_j)(\zeta_{i'}+\zeta_{j'})]^{3/2}} 
e^{-\lambda_{ij}\lvert \mathbf r_i-\mathbf r_j\rvert^2}
e^{-\lambda_{i'j'}\lvert \mathbf r_{i'}-\mathbf r_{j'}\rvert^2} 
\notag\\
&\quad{}\times e^{-P^2/4(\zeta_i+\zeta_j)} e^{-P^2/4(\zeta_{i'}+\zeta_{j'})}\notag\\
&\quad{}\times \sum_{\substack{s^x_{ij},s^y_{ij},s^z_{ij},\\ s^x_{i'j'},s^y_{i'j'},s^z_{i'j'}}}
E_{s^x_{ij}}^{n^x_i n^x_j} 
E_{s^y_{ij}}^{n^y_i n^y_j} 
E_{s^z_{ij}}^{n^z_i n^z_j} 
E_{s^x_{i'j'}}^{n^x_{i'} n^x_{j'}} 
E_{s^y_{i'j'}}^{n^y_{i'} n^y_{j'}} 
E_{s^z_{i'j'}}^{n^z_{i'} n^z_{j'}}
(-1)^{s^x_{i'j'}+s^y_{i'j'}+s^z_{i'j'}} \notag\\
&\quad{}\times 
\left( \frac{\partial}{\partial X_{iji'j'}} \right)^{s^x_{ij}+s^x_{i'j'}} 
\left( \frac{\partial}{\partial Y_{iji'j'}} \right)^{s^y_{ij}+s^y_{i'j'}} 
\left( \frac{\partial}{\partial Z_{iji'j'}} \right)^{s^z_{ij}+s^z_{i'j'}} 
\mathfrak j_0(P,R_{iji'j'}) .
\end{align}
Now we calculate the partial derivatives of $\mathfrak j_n(P,R_{iji'j'})$ with respect to $X_{iji'j'}$, $Y_{iji'j'}$ and $Z_{iji'j'}$.
Using the identity
\begin{align}
\frac{d}{dz}\frac{j_n(z)}{z^n}=- \frac{j_{n+1}(z)}{z^n},
\end{align}
the relation $R_{iji'j'}= \left(X_{iji'j'}^2 + Y_{iji'j'}^2 + Z_{iji'j'}^2\right)^{1/2}$
and the chain rule, we can evaluate the partial derivative of $\mathfrak j_n(P,R_{iji'j'})$ with respect to $X_{iji'j'}$ 
\begin{align}
\frac{\partial \mathfrak j_n(P,R_{iji'j'})}{\partial X_{iji'j'}}
&= \frac{\partial R_{iji'j'}}{\partial X_{iji'j'}}  \frac{\partial \mathfrak j_n(P,R_{iji'j'})}{\partial R_{iji'j'}} ,\notag\\
&= \frac{X_{iji'j'}}{R_{iji'j'}} P^n \frac{\partial}{\partial R_{iji'j'}} \left[\frac{j_n(PR_{iji'j'})}{R_{iji'j'}^n} \right], \notag\\
&= - X_{iji'j'} \left(\frac{P}{R_{iji'j'}}\right)^{n+1} j_{n+1}(PR_{iji'j'}), \notag\\
&= - X_{iji'j'} \mathfrak j_{n+1}(P,R_{iji'j'}). \label{eq:dJdX}
\end{align}
Similarly,
\begin{align}
\frac{\partial \mathfrak j_n(P,R_{iji'j'})}{\partial Y_{iji'j'}}= - Y_{iji'j'} \mathfrak j_{n+1}(P,R_{iji'j'}) , \qquad
\frac{\partial \mathfrak j_n(P,R_{iji'j'})}{\partial Z_{iji'j'}}= - Z_{iji'j'} \mathfrak j_{n+1}(P,R_{iji'j'}). \label{eq:dJdYdZ}
\end{align}

The derivatives required in Equation~\ref{eq:gjffjjnf} can be then be found recursively using the following algorithm. Let 
\begin{equation}\label{eq29}
D_{lmn}(X,Y,Z)= \left( \frac{\partial}{\partial X} \right)^l \left( \frac{\partial}{\partial Y} \right)^m \left( \frac{\partial}{\partial Z} \right)^n \mathfrak j_0,
\end{equation}
where we have simplified the notation slightly by using $X\equiv X_{iji'j'}$, $Y\equiv Y_{iji'j'}$, $Z\equiv Z_{iji'j'}$ and $\mathfrak j_0\equiv \mathfrak j_0(P,R_{iji'j'})$ (which is an implicit function of $X$, $Y$ and $Z$).
From Equations~\ref{eq:dJdX} and \ref{eq:dJdYdZ}, we note that taking a partial derivative of $\mathfrak j_k$ with respect to $X$ gives $-X\mathfrak j_{k+1}$, and similarly for $Y$ and $Z$, and thus, $D_{lmn}(X,Y,Z)$ will be a combination of $\mathfrak j_k$'s, with powers of $X$, $Y$, and $Z$ as coefficients. 
That is, we can write
\begin{align}
D_{lmn}(X,Y,Z) = \sum_{\lambda=0}^l  a^{(l)}_\lambda(X) \sum_{\mu=0}^m  b^{(m)}_\mu(Y) \sum_{\nu=0}^n    c^{(n)}_\nu(Z) \mathfrak j_{\lambda+\mu+\nu}, \label{eq:30}
\end{align}
where $a^{(l)}_\lambda$, $b^{(m)}_\mu$ and $c^{(n)}_\nu$ are polynomials of degree ${\le} l$, $m$ and $n$, respectively.
Without loss of generality, we can determine a recursion relation for the polynomials $a^{(l)}_\lambda$ by considering the case $m=n=0$. In this case, Equation~\ref{eq:30} becomes
\begin{align}
D_{l00}(X,Y,Z) = \sum_{\lambda=0}^l  a^{(l)}_\lambda(X) \mathfrak j_{\lambda}. \label{eq:31}
\end{align}
Differentiating Equation~\ref{eq:31} with respect to $X$ gives
\begin{align}
D_{l+1,00}\equiv\frac{\partial}{\partial X} D_{l00}
&=\frac{\partial}{\partial X}\sum_{\lambda=0}^l  a^{(l)}_\lambda \mathfrak j_{\lambda} 
= \sum_{\lambda=0}^l \left( \frac{\partial a^{(l)}_\lambda}{\partial X}\mathfrak j_\lambda+a^{(l)}_\lambda \frac{\partial\mathfrak j_\lambda}{\partial X}\right),\notag\\
&=\sum_{\lambda=0}^l \left( \frac{\partial a^{(l)}_\lambda}{\partial X}\mathfrak j_\lambda-a^{(l)}_\lambda X \mathfrak j_{\lambda+1}\right). \label{eq:32}
\end{align}
By changing $l\rightarrow l+1$ in Equation~\ref{eq:31} we have a second expression for $D_{l+1,00}$:
\begin{align}
D_{l+1,00} = \sum_{\lambda=0}^{l+1}a^{(l+1)}_\lambda\mathfrak j_\lambda. \label{eq:33}
\end{align}
Setting these two expressions for $D_{l+1,00}$ equal yields
\begin{align}
\sum_{\lambda=0}^{l+1}a^{(l+1)}_\lambda\mathfrak j_\lambda 
- \sum_{\lambda=0}^l  \frac{\partial a^{(l)}_\lambda}{\partial X}\mathfrak j_\lambda
 + \sum_{\lambda=0}^l a^{(l)}_\lambda X \mathfrak j_{\lambda+1} = 0.
\end{align}
Next, we seek to rearrange this equation to obtain recurrence relations for the coefficients. By setting $a^{(l)}_\lambda\equiv 0$ if $\lambda>l$, we may rewrite the second term as a sum up to $l+1$, and 
the third term can be reindexed as $\sum_{\lambda=1}^{l+1} a^{(l)}_{\lambda-1}X\mathfrak j_\lambda$:
\begin{align}
\phantom{\implies} & \sum_{\lambda=0}^{l+1}a^{(l+1)}_\lambda\mathfrak j_\lambda 
- \sum_{\lambda=0}^{l+1}  \frac{\partial a^{(l)}_\lambda}{\partial X}\mathfrak j_\lambda
+ \sum_{\lambda=1}^{l+1} a^{(l)}_{\lambda-1}X\mathfrak j_\lambda = 0, \notag\\
\implies & \left(a^{(l+1)}_0 - \frac{\partial a^{(l)}_0}{\partial X}\right)\mathfrak j_0
+ \sum_{\lambda=1}^{l+1}
\left(a^{(l+1)}_\lambda\mathfrak 
-   \frac{\partial a^{(l)}_\lambda}{\partial X}
+  a^{(l)}_{\lambda-1}X
\right)\mathfrak j_\lambda=0,
\end{align}
where the last expression is obtained by separating the $\lambda=0$ terms from the first two sums.

The functions $\{ \mathfrak j_\lambda \}_{\lambda=0}^{l+1}$ are linearly independent, which leads us to the recurrence relations
\begin{align}\label{eq:rec_rel_a}
a^{(l+1)}_0(X) &= \frac{\partial }{\partial X} a^{(l)}_0(X) , \notag \\
a^{(l+1)}_\lambda(X) &= \frac{\partial }{\partial X}a^{(l)}_\lambda(X) - Xa^{(l)}_{\lambda-1}(X) \qquad(\lambda\ge1),
\end{align}
with $a^{(0)}_0=1$.  The other coefficients are defined similarly:
\begin{align}
b^{(m+1)}_0 &= \frac{\partial b^{(m)}_0}{\partial Y} ,\nonumber \\
b^{(m+1)}_\mu &= \frac{\partial b^{(m)}_\mu}{\partial Y} - b^{(m)}_{\mu-1}Y\qquad(\mu\ge1), \\
c^{(n+1)}_0 &= \frac{\partial c^{(n)}_0}{\partial Z} , \nonumber\\
c^{(n+1)}_\nu &= \frac{\partial c^{(n)}_\nu}{\partial Z} - c^{(n)}_{\nu-1}Z\qquad(\nu\ge1),
\end{align}
with $b^{(0)}_0=c^{(0)}_0=1$. 
Table \ref{tab:coeff} shows expressions for $a^{(l)}_\lambda$ for $l=0$--16, enough for $g$-type functions which are the highest-angular-momentum functions used in {\tt EXCITON+} code. 
The recursion relations in Equation \ref{eq:rec_rel_a} enable each element of the table to be computed easily from the elements  directly above and to the above-left. Corresponding expressions for coefficients $b^{(l)}_\mu$ and $c^{(l)}_\nu$ are obtained in the same way with $X$ replaced by $Y$ and $Z$, respectively. 
\begin{sidewaystable}
\centering
\fontsize{10}{16}\selectfont
\caption{\label{tab:coeff}Expressions for $a^{(l)}_\lambda(X)$ for $l=0$--16. Each element is obtained by taking the derivative with respect to $X$ of the element above, and subtracting $X$ times the element to the above left. 
}
\begin{tabular}{c|ccccccccccccccccccccc}
\hline
$l$ \textbackslash $\lambda$ & 0 & 1 & 2 & 3 & 4 & 5 & 6 & 7 & 8 & 9 & 10  \\
\hline
0 & 1 \\
1 & 0 & $-X$ \\
2 & 0 & $-1$ & $X^2$ \\
3 & 0 & 0 & $3X$ & $-X^3$ \\
4 & 0 & 0 & 3 & $-6X^2$ & $X^4$ \\
5 & 0 & 0 & 0 & $-15X$ & $10X^3$ & $-X^5$ \\
6 & 0 & 0 & 0 & $-15$ & $45X^2$ & $-15X^4$ & $X^6$  \\
7 & 0 & 0 & 0 & 0 & $105X$ & $-105X^3$ & $21X^5$ & $-X^7$   \\
8 & 0 & 0 & 0 & 0 & 105 & $-420X^2$ & $210X^4$ & $-28X^6$ & $X^8$ \\
9 & 0 & 0 & 0 & 0 & 0 & $-945X$ & $1260X^3$ & $-378X^5$ & $36X^7$ & $-X^9$ \\
10 & 0 & 0 & 0 & 0 & 0 & $-945$ & $4725X^2$ & $-3150X^4$ & $630X^6$ & $-45X^8$ & $X^{10}$ \\
\hline
$l$ \textbackslash $\lambda$ & 0--5 & 6 & 7 & 8 & 9 & 10 & 11 & 12 & 13 & 14 & 15 & 16 \\
\hline
11 & 0 & $10395X$ & $-17325X^3$ & $6930X^5$ & $-990X^7$ & $55X^9$ & $-X^{11}$  \\
12 & 0 & 10395 & $-62370X^2$ & $51975X^4$ & $-13860X^6$ & $1485X^8$ & $-66X^{10}$ & $X^{12}$  \\
13 & 0 & 0 & $-135135X$ & $270270X^3$ & $-135135X^5$ & $25740X^7$ & $-2145X^9$ & $78X^{11}$ & $-X^{13}$ \\
14 & 0 & 0 & $-135135$ & $945945X^2$ & $-945945X^4$ & $315315X^6$ & $-45045X^8$ & $3003X^{10}$ & $-91X^{12}$ & $X^{14}$ \\
15 & 0 & 0 & 0 & $2027025X$ & $-4729725X^3$ & $2837835X^5$ & $-675675X^7$ & $75075X^9$ & $-4095X^{11}$ & $105X^{13}$ & $-X^{15}$ \\
16 & 0 & 0 & 0 & $2027025$ & $-16216200X^2$ & $18918900X^4$ & $-7567560X^6$ & $1351350X^8$ & $-120120X^{10}$ & $5460X^{12}$ & $-120X^{14}$ & $X^{16}$ \\
\hline
\end{tabular}
\end{sidewaystable}

Putting all of this together in Equation \ref{eq:gjffjjnf}, we have an expression for $\int \left|A^{(0)}_{n\varepsilon}(\mathbf{P})\right|^2d\Omega_{\mathbf{P}}$:
\begin{align}
    \int \left|A^{(0)}_{n\varepsilon}(\mathbf{P})\right|^2&d\Omega_{\mathbf{P}} 
    =\nonumber
    \sum_{i,j,i',j'}C_i^{\varepsilon} C^{(n)}_j C^{\varepsilon *}_{i'} C^{(n)*}_{j'} 
    e^{-\lambda_{ij}|\mathbf{r}_i-\mathbf{r}_j|^2} e^{-\lambda_{i'j'}|\mathbf{r}_{i'}-\mathbf{r}_{j'}|^2}
    \frac{4\pi^4}{[(\zeta_i+\zeta_j)(\zeta_{i'}+\zeta_{j'})]^{3/2}}\, 
    \\\nonumber
    &\times  e^{-P^2/4(\zeta_i+\zeta_j)} e^{-P^2/4(\zeta_{i'}+\zeta_{j'})}  \\\nonumber
&\times \sum_{\substack{s^x_{ij},s^y_{ij},s^z_{ij},\\ s^x_{i'j'},s^y_{i'j'},s^z_{i'j'}}}
E_{s^x_{ij}}^{n^x_i n^x_j} 
E_{s^y_{ij}}^{n^y_i n^y_j} 
E_{s^z_{ij}}^{n^z_i n^z_j} 
E_{s^x_{i'j'}}^{n^x_{i'} n^x_{j'}} 
E_{s^y_{i'j'}}^{n^y_{i'} n^y_{j'}} 
E_{s^z_{i'j'}}^{n^z_{i'} n^z_{j'}}
(-1)^{s^x_{i'j'}+s^y_{i'j'}+s^z_{i'j'}}
   \\ 
&\times 
\!\!\!\sum_{\lambda=0}^{s^x_{ij}+s^x_{i'j'}} \!\!\!a^{(s^x_{ij}+s^x_{i'j'})}_\lambda(X_{iji'j'})
\!\!\!\sum_{\mu=0}^{s^y_{ij}+s^y_{i'j'}} \!\!\!b^{(s^y_{ij}+s^y_{i'j'})}_\mu(Y_{iji'j'}) 
\!\!\!\sum_{\nu=0}^{s^z_{ij}+s^z_{i'j'}}   \!\!\!c^{({s^z_{ij}+s^z_{i'j'}})}_\nu(Z_{iji'j'}) 
\mathfrak j_{\lambda+\mu+\nu}(P,R_{iji'j'}).
\label{eq:ann_amp_int_final}\end{align}

Here, we have explicitly written the complex conjugates of the expansion coefficients $C^{\varepsilon *}_{i'}$ and $C^{(n)*}_{j'}$, but in practice our basis functions are real and the corresponding expansion coefficients are real as well so the complex conjugate is not required. We now turn our attention to the final integral over $P$ in Equation~\ref{eqn:gamspec}.
Substituting Equation \ref{eq:ann_amp_int_final} into Equation~\ref{eqn:gamspec}, we have
\begin{equation}
w^{(0)}_{n \varepsilon}(\epsilon) = 
\sum_{i,i',j,j'}
C^{\varepsilon}_{i}
C^{\varepsilon}_{i'}
C^{n}_{j}
C^{n}_{j'}
[\gamma(\epsilon)]_{i',j'}^{i,j},
\label{eq:gamma_expansion}
\end{equation}
where we introduce the \emph{$\gamma$-spectrum matrix}
\begin{align}
[\gamma(\epsilon)]_{i',j'}^{i,j}&= 
\frac{1}{(2\pi)^3c} 
\frac{4\pi^4}{[(\zeta_i+\zeta_j)(\zeta_{i'}+\zeta_{j'})]^{3/2}} 
e^{-\lambda_{ij}\lvert \mathbf r_i-\mathbf r_j\rvert^2}
e^{-\lambda_{i'j'}\lvert \mathbf r_{i'}-\mathbf r_{j'}\rvert^2} 
\notag\\
&\quad{}\times \sum_{\substack{s^x_{ij},s^y_{ij},s^z_{ij},\\ s^x_{i'j'},s^y_{i'j'},s^z_{i'j'}}}
E_{s^x_{ij}}^{n^x_i n^x_j} 
E_{s^y_{ij}}^{n^y_i n^y_j} 
E_{s^z_{ij}}^{n^z_i n^z_j} 
E_{s^x_{i'j'}}^{n^x_{i'} n^x_{j'}} 
E_{s^y_{i'j'}}^{n^y_{i'} n^y_{j'}} 
E_{s^z_{i'j'}}^{n^z_{i'} n^z_{j'}}
(-1)^{s^x_{i'j'}+s^y_{i'j'}+s^z_{i'j'}} \notag\\
&\quad{} \times \sum_{\lambda=0}^{s^x_{ij}+s^x_{i'j'}}a^{(s^x_{ij}+s^x_{i'j'})}_\lambda (X_{iji'j'})
 \sum_{\mu=0}^{s^y_{ij}+s^y_{i'j'}} b^{(s^y_{ij}+s^y_{i'j'})}_\mu  (Y_{iji'j'})
 \sum_{
 \nu=0}^{s^z_{ij}+s^z_{i'j'}}   c^{(s^z_{ij}+s^z_{i'j'})}_{\nu} (Z_{iji'j'})
\mathfrak I^{(\lambda\mu \nu)}_{iji'j'}(\epsilon) ,
\end{align}
where
\begin{align}
\mathfrak I^{(\lambda\mu \nu)}_{iji'j'}(\epsilon) = 
\int_{2\lvert\epsilon\rvert/c}^\infty P \exp\left[-\frac{P^2}{4}\left(\frac{1}{\zeta_i+\zeta_j}+\frac{1}{\zeta_{i'}+\zeta_{j'}}\right)\right] \mathfrak j_{\lambda+\mu+\nu}(P,R_{iji'j'})\,dP.\label{eq53}
\end{align}
The integral in Equation~\ref{eq53} is computed numerically using Gauss-Legendre quadrature. To do this, an appropriate finite upper limit for the integral, $P_{\text{max}}$, must be chosen based on the decay of the integrand as $P\rightarrow\infty$. The long-range behaviour of the integrand is dominated by the decaying Gaussian factor, so $P_\text{max}$ is chosen such that
$P_\text{max}^2[1/(\zeta_i+\zeta_j)+1/(\zeta_{i'}+\zeta_{j'})]/4\gg1$, i.e., $P_\text{max}\gg2/[1/(\zeta_i+\zeta_j)+1/(\zeta_{i'}+\zeta_{j'})]^{1/2}$. In practice, we use the limit
\begin{equation}
P_\text{max}=10/[1/(\zeta_i+\zeta_j)+1/(\zeta_{i'}+\zeta_{j'})]^{1/2}.
\end{equation}

\end{document}